\documentclass[12pt]{article}

\usepackage{graphics}
\usepackage{epsfig}
\usepackage{cite}
\input epsf

\title{Heavy Flavour Production at Tevatron and Parton Shower Effects}

\author{H.~Jung$^{1,2}$, M.~Kraemer$^1$, A.V.~Lipatov$^3$, N.P.~Zotov$^3$}

\begin{document}
\maketitle

\vspace*{-7.5cm}
\begin{flushright}
DESY 10-134\\
Dec 2010 \\
version 2(8)
\end{flushright}
\vspace*{+4.5cm}

\begin{center}

{\it $^1$DESY, Hamburg, Germany\\[3mm]
$^2$University of Antwerp, Antwerp, Belgium\\[3mm]
$^3$D.V.~Skobeltsyn Institute of Nuclear Physics,\\ 
M.V. Lomonosov Moscow State University, Russia}\\[3mm]

\end{center}

\vspace{0.5cm}

\begin{center}

{\bf Abstract }

\end{center}

We present hadron-level predictions from the Monte Carlo generator {\sc Cascade} and 
numerical calculations of charm and beauty production 
at the Fermilab Tevatron within the framework of the $k_T$-factorization QCD approach.
Our consideration is based on the CCFM-evolved unintegrated gluon densities in a proton.
The performed analysis covers the total and differential cross sections 
of open charm and beauty quarks, $B$ and $D$ mesons (or rather muons from their 
semileptonic decays) and the total and differential cross sections of 
$b \bar b$ di-jet hadroproduction. We study the theoretical uncertainties
of our calculations and investigate the effects coming from parton 
showers in initial and final states.
Our predictions are compared with the recent
experimental data taken by the D0 and CDF collaborations. 
Special attention is put on the specific angular correlations between the 
final-state particles. 
We demonstrate that the final state parton shower plays a crucial role in the description of 
such observables. The decorrelated part of angular separations can be fully 
described, if the process $gg^*\rightarrow gg$ is included. 

\vspace{0.8cm}

\noindent
PACS number(s): 12.38.-t, 13.85.-t

\vspace{0.5cm}
\newpage

\section{Introduction} \indent 

Charm and beauty production at high energies is subject of intense 
studies from both theoretical and experimental 
points of view~\cite{1,2,3,4,5,6}. From the theoretical point, these 
processes provide an opportunity to test the different 
predictions based on Quantum Chromodynamics (QCD)
since the dominant production mechanism at high energies 
(i.e. small $x$) is believed
to be quark pair production through the gluon-gluon fusion subprocess.
At present, the problem of description of the charm and beauty production at the Tevatron
within QCD is not fully solved. So, the difference between the D0 and CDF 
measurements~\cite{1,2,3,4} of the $b$-quark and $B$-meson production cross sections
and the calculations~\cite{7} performed in the framework of fixed-order 
next-to-leading logarithm scheme (FONLL) is about of a factor of~1.7.
The central FONLL predictions~[8] also lie below the data on 
the $D^0$, $D^+$, $D^{*+}$ and $D^{+}_s$ cross sections
which has been measured~[5] by the CDF collaboration 
as a functions of their transverse momenta.
Recently the calculations
in the general-mass variable-flavour-number scheme (GM-VFNS)
were performed~\cite{9,10} for the transverse momentum distribution of 
$B$ and $D$ mesons and have been found to be consistent with the data~\cite{4, 5}.
However, there is still no GM-VFNS predictions for other measured quantities
(like distributions in rapidity and azimuthal angle difference between the momenta
of final mesons).

Heavy flavour production
has been considered~\cite{11,12,13,14,15,16} also in the framework of the $k_T$-factorization 
QCD approach~\cite{17}. 
This approach is based on the Balitsky-Fadin-Kuraev-Lipatov (BFKL)~[18] or 
Ciafaloni-Catani-Fiorani-Marchesini (CCFM)~\cite{19} equations for the non-collinear 
gluon evolution in a proton and gives the possibility to take into account
large logarithmic terms (proportional to $\ln 1/x$ and, in the case of CCFM, 
also to $\ln 1/(1 - x)$).
A detailed description and discussion of the $k_T$-factorization approach can be
found in~[20]. A reasonable agreement between the
$k_T$-factorization predictions and the Tevatron data on the heavy flavour production
has been found in~\cite{12,13,14,15,16}. It was demonstrated~\cite{11, 15, 16} that
studying the specific angular correlations between the transverse
momenta of produced quarks can give an unique information about 
the non-collinear gluon evolution in a proton since taking 
into account the non-vanishing initial gluon
transverse momentum ${\mathbf k}_{T}$ in the $k_T$-factorization approach leads to 
the violation of back-to-back kinematics event at leading order.

Measurements of $b \bar b$ di-jet total and differential 
cross sections (as a functions of the leading jet transverse energy $E_T$ and 
the di-jet invariant mass) and azimuthal angle correlations between two $b$-jets
have been performed by the CDF collaboration~\cite{21}. These measurements significantly 
extend the energy range investigated by previous analyses~\cite{1,2,3,4,5}.
Studying the $b$-jet production is specially interesting since there are
no additional assumptions on the fragmentation of the beauty quark into the $B$ meson.
The CDF collaboration has reported the preliminary
data~\cite{6} on the charm pair production, where the $D^0$, $D^{*-}$ 
pair cross section and the $D^+$, $D^{*-}$ pair cross section as a function
of the azimuthal angle between two charmed mesons have been meausred.
The angular correlations in $B$-meson production have been also measured~\cite{1,2,3}
by the D0 and CDF collaborations. 

The main goal of present paper is to give a systematic analysis of all available 
experimental data~\cite{1,2,3,4,5,6,21} on the heavy flavour production at the Tevatron in the
framework of the $k_T$-factorization formalism. We produce the relevant 
numerical calculations in two ways. First, we will perform analytical 
parton-level calculations (labeled as LZ in the following) 
similar to that  done in~\cite{15,16}. In these calculations we will 
use the CCFM-evolved gluon densities~\cite{22} as default sets.
The measured cross sections of heavy quark production will be compared
with the predictions of Monte Carlo event generator 
\textsc{Cascade}~\cite{23}. \textsc{Cascade} is a full hadron level Monte Carlo event 
generator for $ep$, $\gamma p$, $p\bar p$ and $pp$ processes, which uses the CCFM 
evolution equation for the initial state cascade in
a backward evolution approach supplemented with off-shell matrix elements for the
hard scattering\footnote{A discussion of the phenomenological applications of \textsc{Cascade} 
can be found in~\protect\cite{24}.}.
In this way we will investigate the influence of parton showers in 
initial and final states for the description of the data. 
We will study the possible sources of theoretical uncertainties 
of our predictions (i.e. uncertainties connected with the gluon evolution scheme, 
heavy quark mass, hard scale 
of partonic subprocess and the heavy quark fragmentation functions). 
To investigate the dependence of our predictions on 
the non-collinear evolution scheme we will apply the unintegrated gluon densities derived 
from the usual (DGLAP-evolved) parton distributions (in the Kimber-Martin-Ryskin (KMR)~\cite{25} 
approximation). Our special goal is to study specific kinematic properties of the final 
heavy quark-antiquark pair which are strongly related to the non-zero initial gluon 
transverse momentum.

The outline of our paper is following. In Section~2 we 
recall shortly the basic formulas of the $k_T$-factorization approach with a brief 
review of calculation steps. In Section~3 we present the numerical results
of our calculations and a discussion. Section~4 contains our conclusions.

\section{Theoretical framework} \indent 

The main formulas have been obtained previously in~\cite{14,15,16}. Here we only recall some of them.
The cross section of heavy quark hadroproduction 
at high energies in the $k_T$-factorization approach
is calculated as a convolution of the off-shell (i.e. $k_T$-dependent)
partonic cross section $\hat \sigma$ and the unintegrated gluon 
distributions in a proton. It can be presented in the following form:
$$
  \displaystyle \sigma (p \bar p \to Q\bar Q \, X) = \int {1\over 16\pi (x_1 x_2 s)^2 } {\cal A}(x_1,{\mathbf k}_{1T}^2,\mu^2) {\cal A}(x_2,{\mathbf k}_{2T}^2,\mu^2) |\bar {\cal M}(g^* g^* \to Q\bar Q)|^2 \times \atop
  \displaystyle  \times d{\mathbf p}_{1T}^2 d{\mathbf k}_{1T}^2 d{\mathbf k}_{2T}^2 dy_1 dy_2 {d\phi_1 \over 2\pi} {d\phi_2 \over 2\pi}, \eqno (1)
$$

\noindent 
where ${\cal A}(x,{\mathbf k}_{T}^2,\mu^2)$ is the
unintegrated gluon distribution in a proton, 
$|\bar {\cal M}(g^* g^* \to Q\bar Q)|^2$ is the 
off-shell (i.e. depending on the initial gluon virtualities 
${\mathbf k}_{1T}^2$ and ${\mathbf k}_{2T}^2$) matrix element squared 
and averaged over initial gluon 
polarizations and colors, and 
$s$ is the total center-of-mass energy.
The produced heavy quark $Q$ and anti-quark $\bar Q$ have the 
transverse momenta ${\mathbf p}_{1T}$ and
${\mathbf p}_{2T}$ and the center-of-mass rapidities $y_1$ and $y_2$.
The initial off-shell gluons have a fraction $x_1$ and $x_2$ 
of the parent protons longitudinal 
momenta, non-zero transverse momenta ${\mathbf k}_{1T}$ and 
${\mathbf k}_{2T}$ (${\mathbf k}_{1T}^2 = - k_{1T}^2 \neq 0$, 
${\mathbf k}_{2T}^2 = - k_{2T}^2 \neq 0$) and azimuthal angles
 $\phi_1$ and $\phi_2$. 
The analytic expression for the 
$|\bar {\cal M}(g^* g^* \to Q\bar Q)|^2$ can be found, for example, in~\cite{14, 17}.

The unintegrated gluon distribution in a 
proton ${\cal A}(x,{\mathbf k}_{T}^2,\mu^2)$ in (1)
can be obtained from the analytical or numerical solution of the BFKL or CCFM
evolution equations. In the numerical calculations we have tested 
a few different sets. First of them (A0) was obtained in~\cite{22} 
from the CCFM evolution equation. The initial (starting) distribution
${\cal A}_0(x,{\mathbf k}_{T}^2,\mu_0^2)$ has been used in 
the following form:
$$
  x {\cal A}_0(x,{\mathbf k}_{T}^2,\mu_0^2) = N x^{p_0} (1 - x)^{p_1}\, \exp(-{\mathbf k}_{T}^2/k_0^2), \eqno(2)
$$

\noindent
where all input parameters have been fitted to describe the proton structure function $F_2(x, Q^2)$.
An equally good fit can be obtained using different values for the soft cut 
and a different value for the width of the intrinsic ${\mathbf k}_{T}$ distribution 
(the CCFM set B0). A reasonable description of the $F_2$ data
can be achieved~\cite{22} by both these sets.

To evaluate the unintegrated gluon densities in a proton ${\cal A}(x,{\mathbf k}_T^2,\mu^2)$ 
we apply also the KMR approach~\cite{25}. The KMR approach is a formalism to construct the 
unintegrated parton (quark and gluon) distributions from the known conventional parton
distributions $xa(x,\mu^2)$, where $a = g$ or $a = q$. 
In this scheme, the unintegrated gluon distribution is given by the expression
$$
  \displaystyle {\cal A}(x,{\mathbf k}_T^2,\mu^2) = T_g({\mathbf k}_T^2,\mu^2) {\alpha_s({\mathbf k}_T^2)\over 2\pi} \times \atop {
  \displaystyle \times \int\limits_x^1 dz \left[\sum_q P_{gq}(z) {x\over z} q\left({x\over z},{\mathbf k}_T^2\right) + P_{gg}(z) {x\over z} g\left({x\over z},{\mathbf k}_T^2\right)\Theta\left(\Delta - z\right) \right],} \eqno (3)
$$

\noindent
where $P_{ab}(z)$ are the usual unregulated leading order DGLAP splitting 
functions, $q(x,\mu^2)$ and $g(x,\mu^2)$ are the conventional quark 
and gluon densities\footnote{Numerically, we have used the 
standard GRV 94~(LO)~\cite{26}, MSTW 2008 (LO)~\cite{27} (in LZ calculations) and 
MRST 99~\cite{28} (\textsc{Cascade}) sets.} 
and $T_g({\mathbf k}_T^2,\mu^2)$ is 
the Sudakov form factor. The theta function $\Theta(\Delta - z)$ implies the angular-ordering constraint 
$\Delta = \mu/(\mu + |{\mathbf k}_T|)$ specifically to the last evolution step to 
regulate the soft gluon singularities. 

In Fig.~\ref{fig1} we plot the CCFM set A0 (as a solid lines), the CCFM set B0 (as a dashed lines) 
and KMR (as a dotted lines) unintegrated gluon densities at probing scale
$\mu^2 = 100$~GeV$^2$ as a function of ${\mathbf k}_T^2$ for different values of $x$.
One can see that the shapes of gluon densities under consideration
are very different from each other
In the following we will study the possible manifestations
in the total and differential heavy flavour cross sections.

In our analytical calculations (LZ) the multidimensional integrations in eq.(1)
have been performed
by the means of Monte Carlo technique, using the routine \textsc{vegas}~\cite{29}.
The full C$++$ code is available from the authors on 
request\footnote{lipatov@theory.sinp.msu.ru}.

\section{Numerical results} \indent

We now are in a position to present our numerical results. First we describe our
input and the kinematic conditions. After we fixed the unintegrated
gluon distributions, the cross section (1) depends on
the renormalization and factorization scales $\mu_R$ and $\mu_F$. In the numerical calculations we set 
$\mu_R^2 = m_Q^2 + ({\mathbf p}_{1T}^2 + {\mathbf p}_{2T}^2)/2$,
$\mu_F^2 = \hat s + {\mathbf Q}_T^2$ (where ${\mathbf Q}_T$ is the 
transverse momentum of initial off-shell gluon pair),
$m_c = 1.4 \pm 0.1$~GeV, $m_b = 4.75 \pm 0.25$~GeV and use the LO formula 
for the coupling constant $\alpha_s(\mu^2)$ with $n_f = 4$ active quark flavours
at $\Lambda_{\rm QCD} = 200$~MeV, such 
that $\alpha_s(M_Z^2) = 0.1232$. 

\subsection{Inclusive charm and beauty production} \indent

We begin the discussion by studying the role of non-zero gluon transverse 
momentum $k_T$ in the off-shell matrix elements involved in~(1). 
In Fig.~\ref{fig2} we plot the differential
cross section for $c\bar c$ and $b\bar b$ pair production as a function of 
transverse momentum ${\mathbf p}_T^Q$, rapidity $y^Q$ and the 
azimuthal angle difference between the transverse 
momenta of produced quarks $\Delta \phi^{QQ}$ at $\sqrt s = 1960$~GeV.
The solid histograms correspond to the results obtained according to the 
master formula in eq.(1). The dotted histograms are obtained by using the same formula
but without virtualities of the incoming gluons in partonic amplitude
and with the additional requirement ${\mathbf k}_{1,2 \,T}^2 < \mu_R^2$.
There are no cuts applied on the phase space of the produced quarks.
As it was expected, in the back-to-back region $\Delta \phi^{QQ} \sim \pi$ both results 
coincide with each other. However, we find that a 
sizeable effect appears at low $\Delta \phi^{QQ}$. Therefore the non-zero 
gluon transverse momentum
in the hard matrix element is important for the description of the data
at low and mediate $\Delta \phi^{QQ}$. This effect is more significant for charm production
due to smaller $x$.
We have checked  that the predictions between the LZ 
and \textsc{Cascade} calculations agree well at parton level.

\begin{figure}
\centering
\begin{picture}(16.5,15.)(0.,0.)
\put(-0.6,8.55){\epsfig{figure=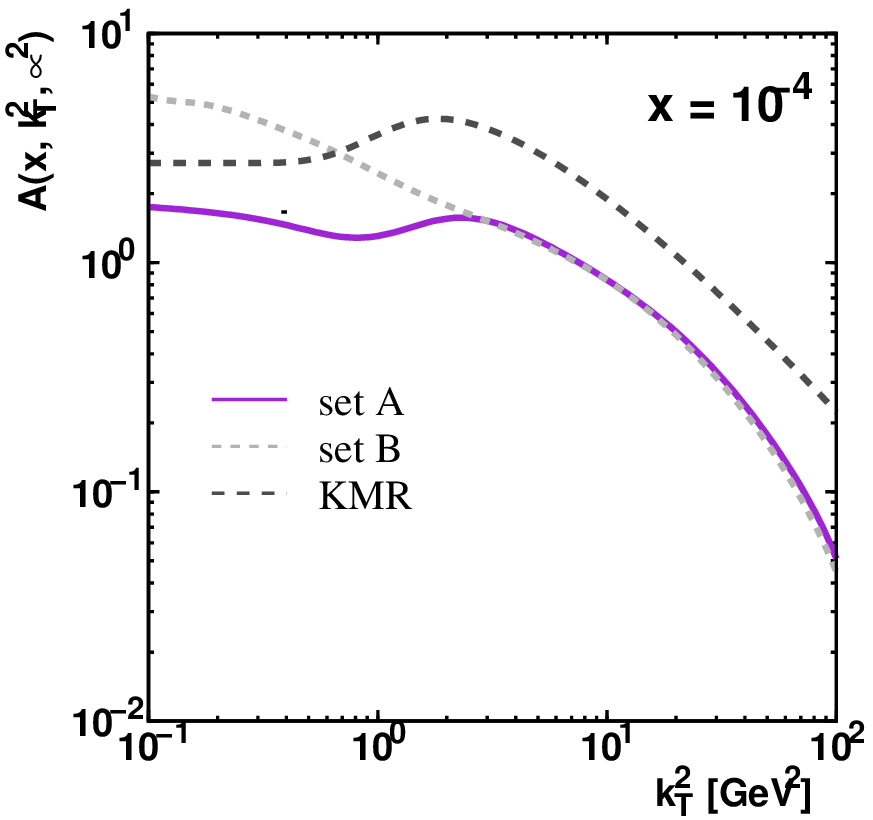, height = 7.4cm, width =8.5cm}}
\put(8.0,8.55){\epsfig{figure=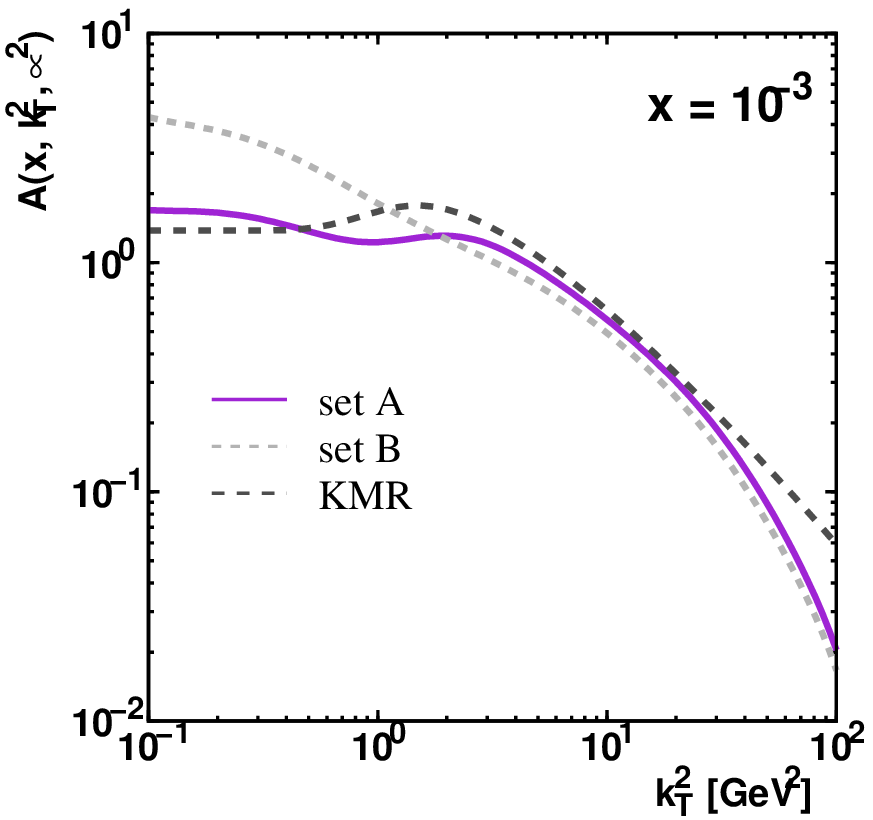, height = 7.4cm, width = 8.5cm}}
\put(-0.6,0.5){\epsfig{figure=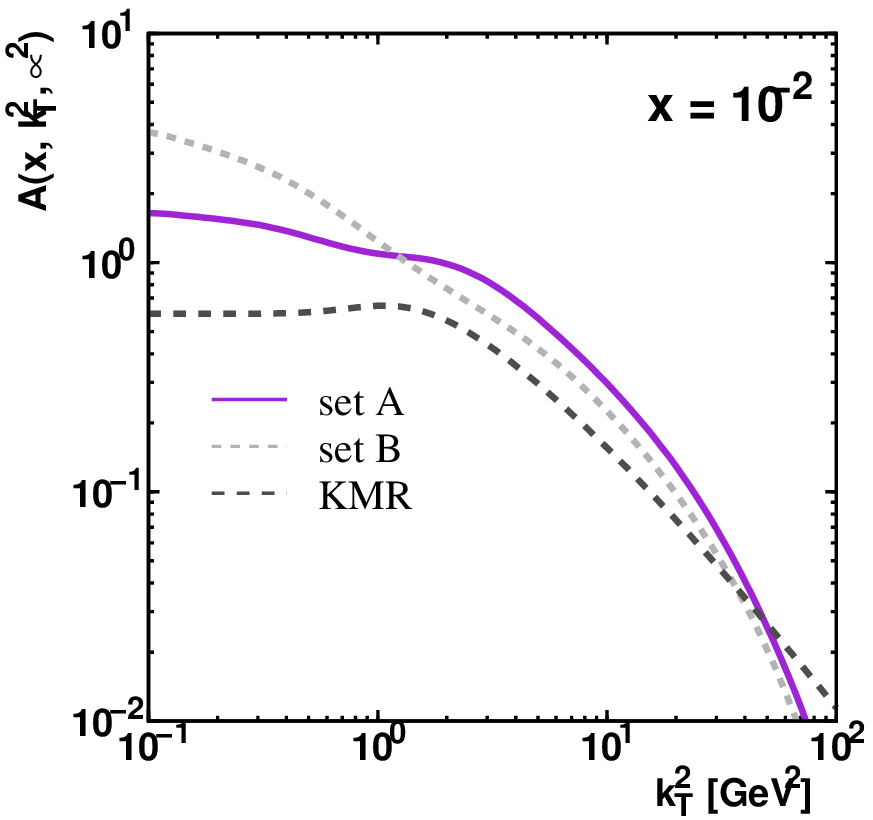, height = 7.4cm, width = 8.5cm}}
\put(8.0,0.5){\epsfig{figure=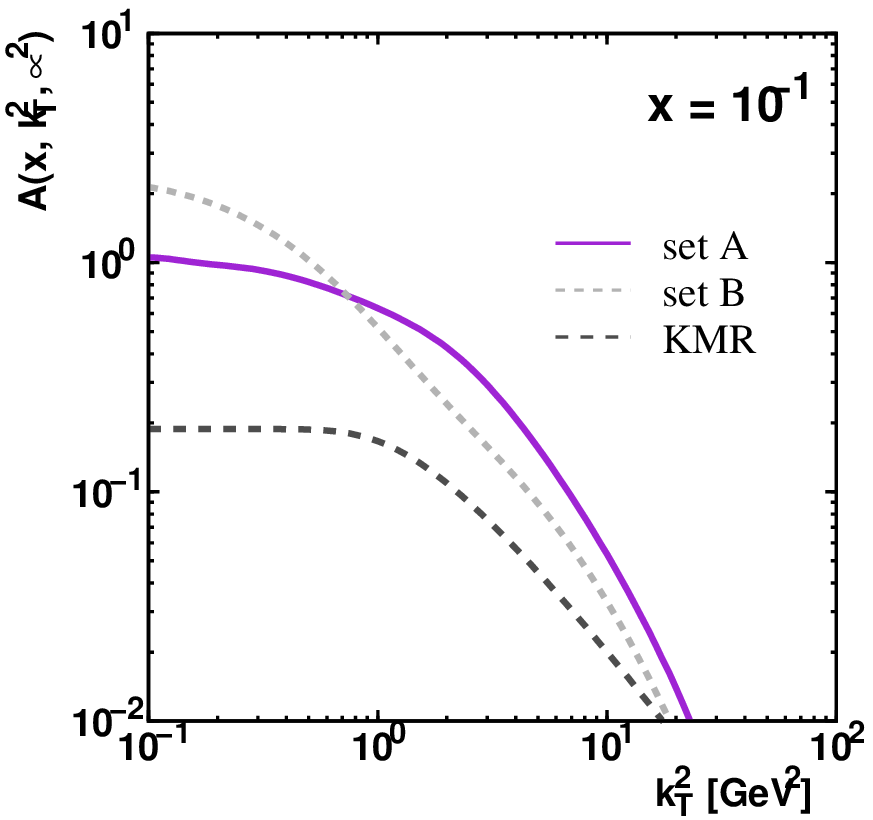, height = 7.4cm, width = 8.5cm}}
\put(4.3,16.){(a)}
\put(12.8,16.){(b)}
\put(4.3,8.){(c)}
\put(12.8,8.){(d)}
\end{picture}
\caption{The unintegrated gluon densities in a proton ${\cal A}(x,{\mathbf k}_{T}^2,\mu^2)$ 
as a function of ${\mathbf k}_{T}^2$ at $\mu^2 = 100$~GeV$^2$.
The solid, dashed and dotted curves 
correspond to the CCFM set A0, CCFM set B0 and KMR distributions, respectively.}
\label{fig1}
\end{figure}

\begin{figure}
\centering
\begin{picture}(16.5,15.)(0.,0.)
\put(-2.2,8.55){\epsfig{figure=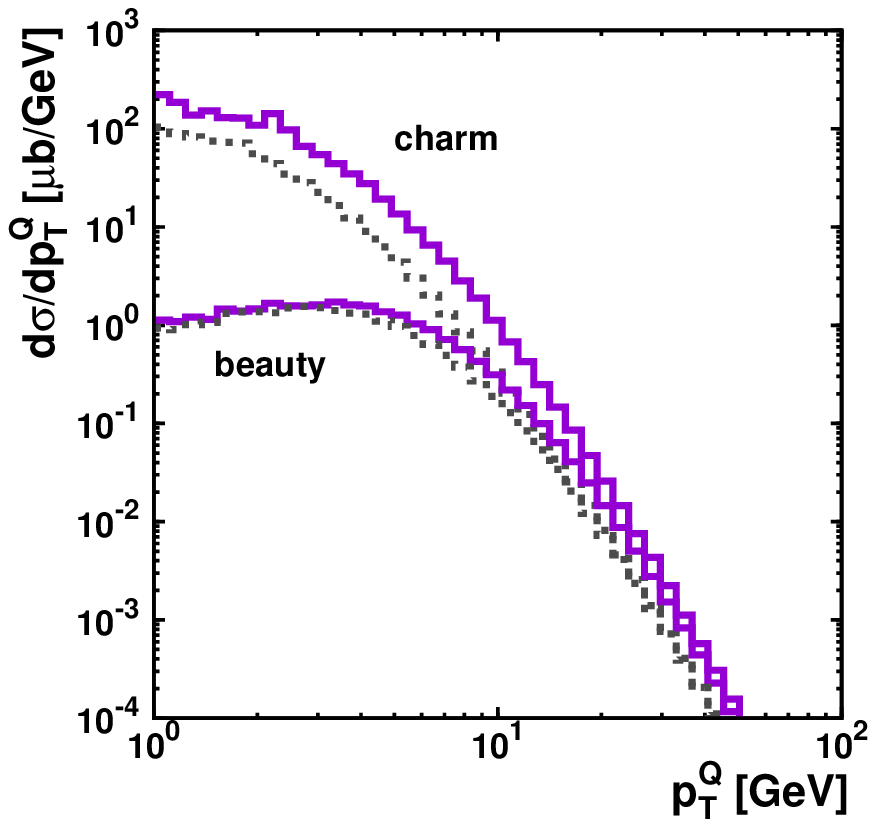, height = 7.8cm, width = 12.2cm}}
\put(6.4,8.55){\epsfig{figure=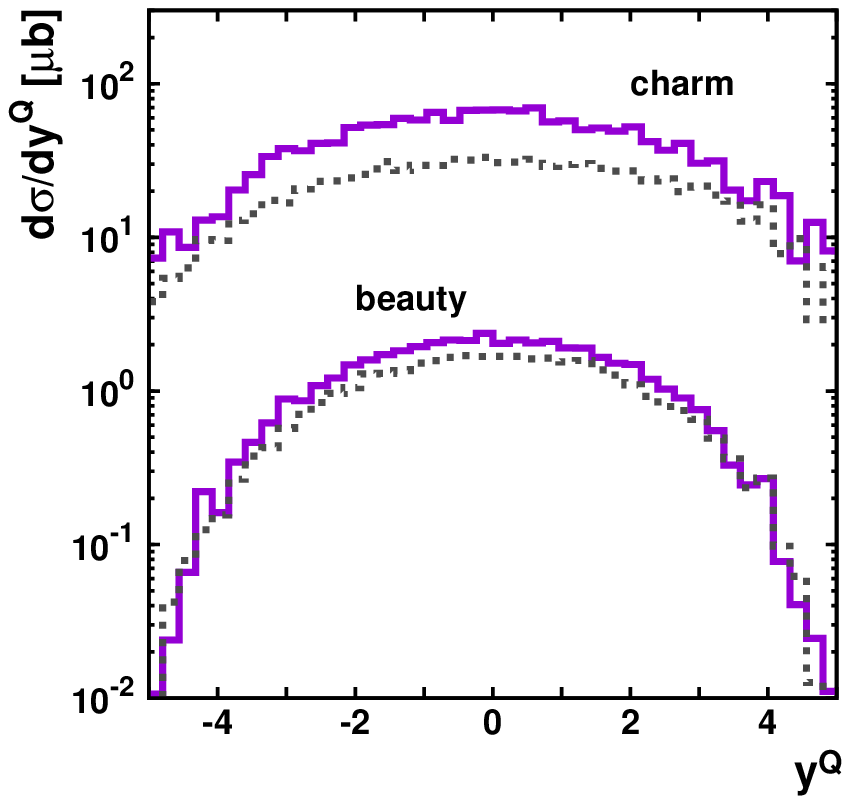, height = 7.8cm, width = 12.2cm}}
\put(2.2,0.5){\epsfig{figure=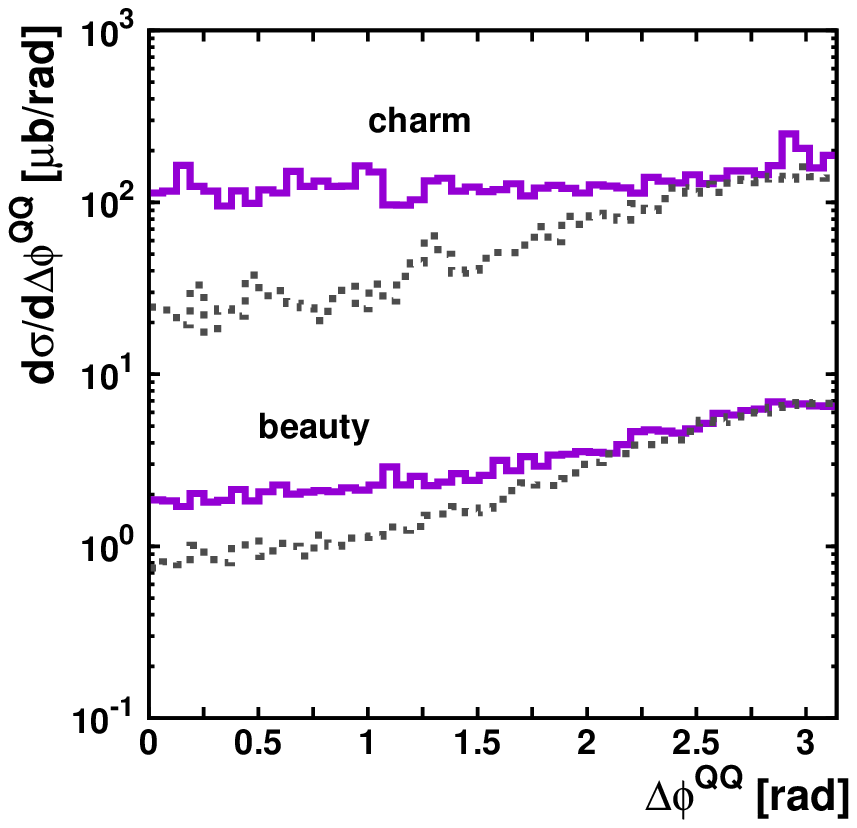, height = 7.8cm, width = 12.2cm}}
\put(4.3,16.){(a)}
\put(12.8,16.){(b)}
\put(8.7,8.){(c)}
\end{picture}
\caption{Importance of non-zero transverse momentum of incoming gluons in charm and 
beauty production at the Tevatron. The solid histograms correspond to the results 
obtained according to the master formula~(1). The dotted histograms are obtained by using 
the same formula but now we switch off the virtualities of both incoming gluons in 
partonic amplitude and apply an additional requirement ${\mathbf k}_{1,2 \,T}^2 < \mu_R^2$.}
\label{fig2}
\end{figure}

\begin{figure}
\centering
\begin{picture}(16.5,15.)(0.,0.)
\put(-2.2,8.55){\epsfig{figure=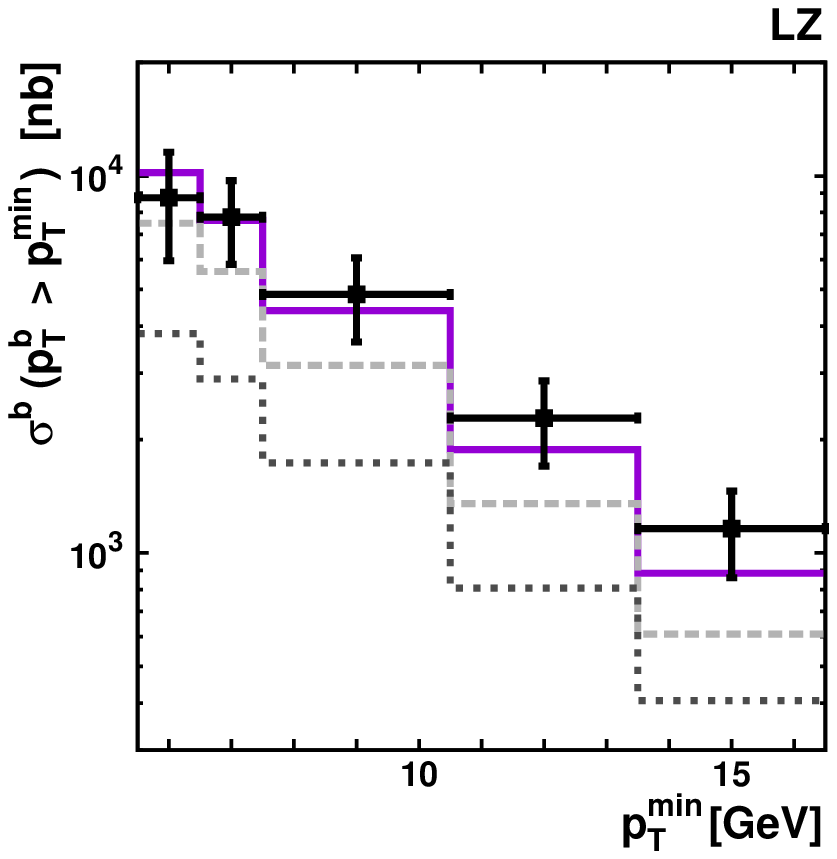, height = 7.8cm, width = 12.2cm}}
\put(7.9,8.){\epsfig{figure=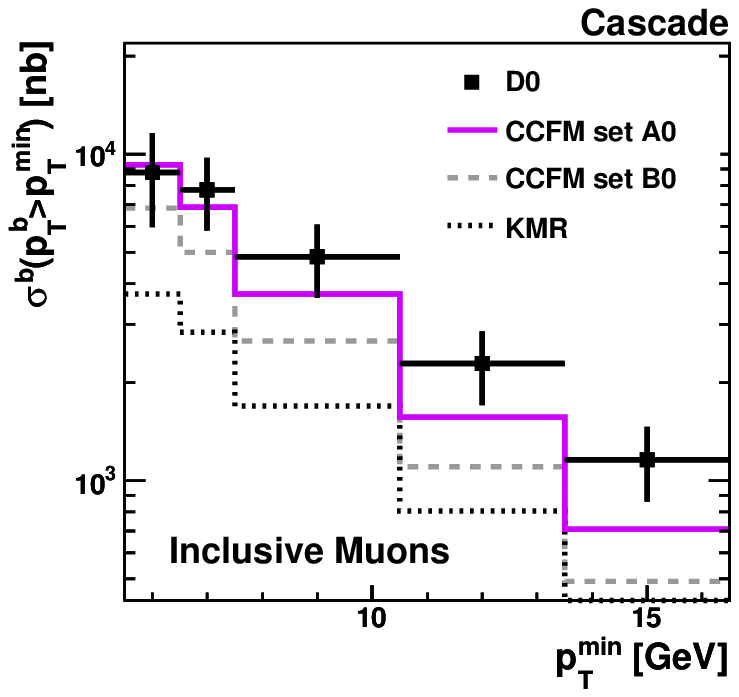, width = 10.4cm}}
\put(-2.2,0.5){\epsfig{figure=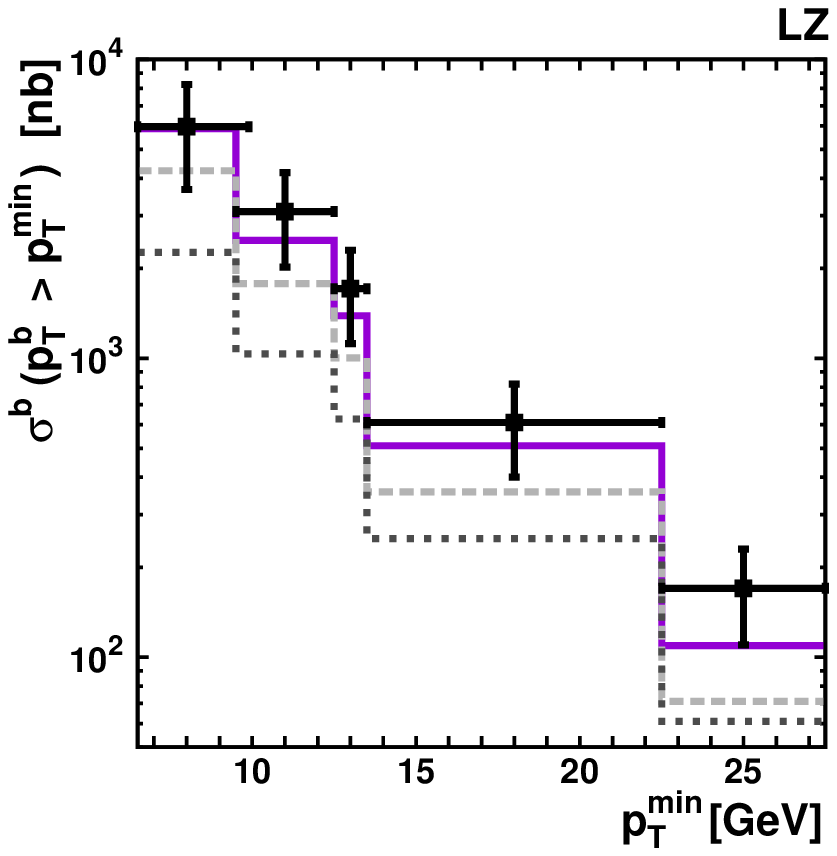, height = 7.8cm, width = 12.2cm}}
\put(7.9,0.0){\epsfig{figure=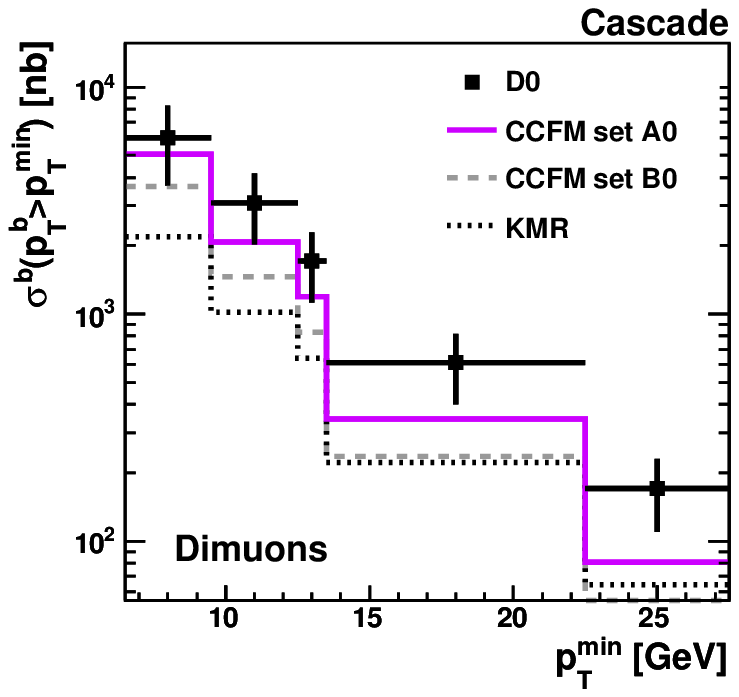, width = 10.4cm}}
\put(4.3,16.){(a)}
\put(12.8,16.){(b)}
\put(4.3,8.){(c)}
\put(12.8,8.){(d)}
\end{picture}
\caption{The transverse momentum distributions of open $b$-quark production at 
the Teavtron. The kinematical cuts applied are described in the text. 
The solid, dashed and dotted histograms 
correspond to the results obtained with the CCFM set A0, CCFM set B0
and KMR unintegrated gluon densities.
The first column shows the LZ results while the second one 
depicts the \textsc{Cascade} predictions. The experimental data are from D0~\protect\cite{1}.}
\label{fig3}
\end{figure}

Now we turn to the transverse momentum distributions of charm and beauty
production. In the case of inclusive $b$-quark production,
the transverse momentum distribution has been measured~\cite{1} by the D0 collaboration and 
has been presented in form of integrated cross section (as a 
function of $b$-quark minimal transverse momentum $p_{T\,{\rm min}}^b$)
 at the total $p\bar p$ energy 
$\sqrt s = 1800$~GeV for  $|y^b| < 1$ 
(note that there is no cut on the rapidity of $\bar b$).
Our predictions are shown in Fig.~\ref{fig3} and compared to the data.
We find good agreement 
between the LZ and \textsc{Cascade} predictions.
We find a good description of the data when using
 the CCFM-evolved (A0) gluon distribution.
The results obtained by using the KMR gluon density are rather close
to the NLO pQCD ones~\cite{30} (not shown) but lie  below the data.
The difference between the CCFM and KMR predictions comes from
 the different behaviour of these gluon densities (see Fig.~\ref{fig1}) which is due to
absence of small-$x$ resummation in the KMR distributions.

\begin{figure}
\centering
\begin{picture}(16.5,8.)(0.,0.)
\put(-2.3,0.6){\epsfig{figure=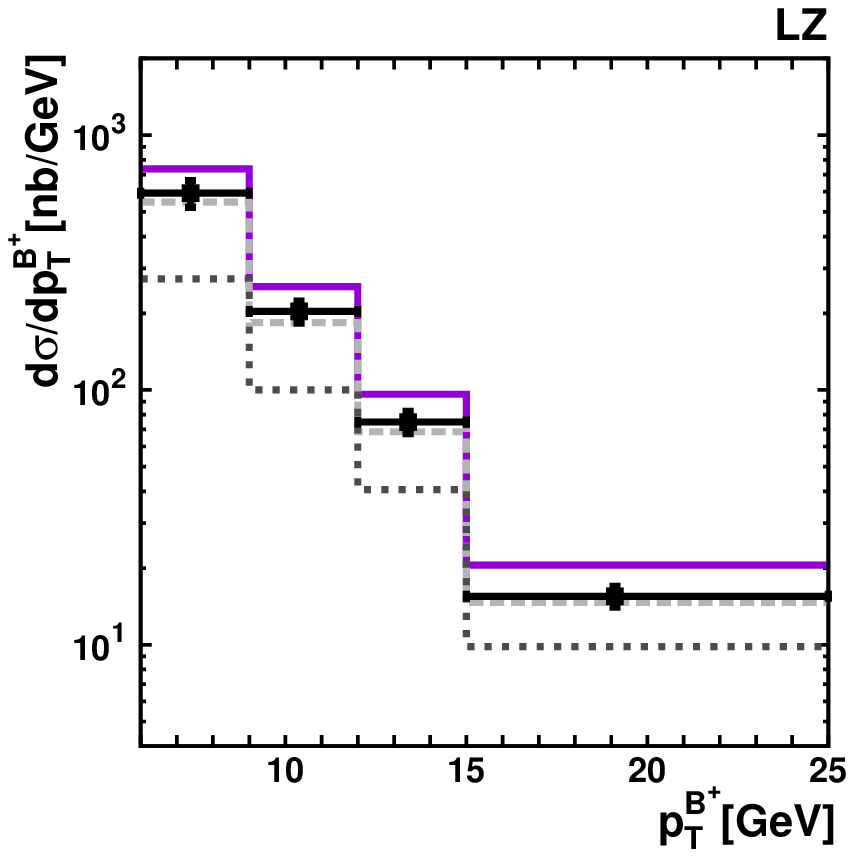, height = 7.7cm, width = 12.2cm}}
\put(7.9,0.0){\epsfig{figure=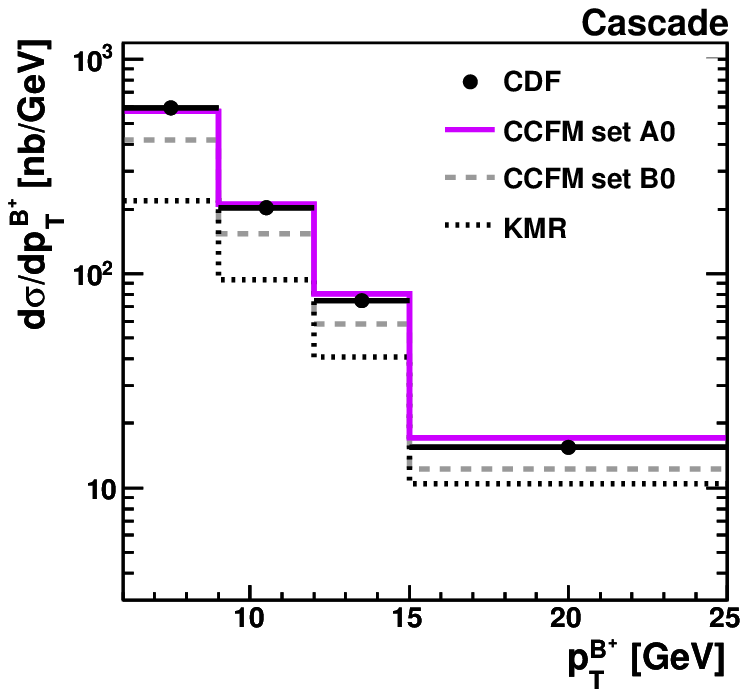, width = 10.4cm}}
\put(4.3,8.){(a)}
\put(12.8,8.){(b)}
\end{picture}
\caption{The cross section as a function of transverse momentum of 
$B^+$ meson hadroproduction. The kinematical cuts applied are described in the text. 
The left histogram shows the LZ numerical 
results while the right plot depicts the \textsc{Cascade} predictions.
Notation of all histograms is the same as in Fig.~\protect\ref{fig3}.
The experimental data are from CDF~\protect\cite{4}.}
\label{fig4}
\end{figure}

\begin{figure}
\centering
\begin{picture}(16.5,15.)(0.,0.)
\put(-2.3,8.55){\epsfig{figure=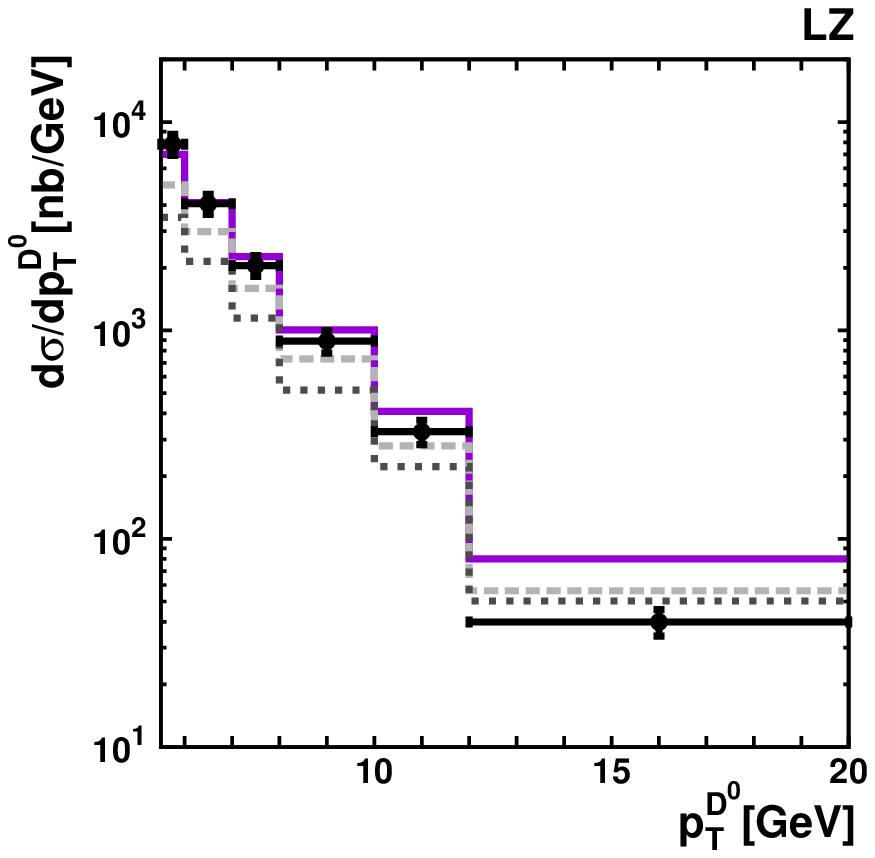, height = 7.8cm, width = 12.2cm}}
\put(8.0,8.){\epsfig{figure=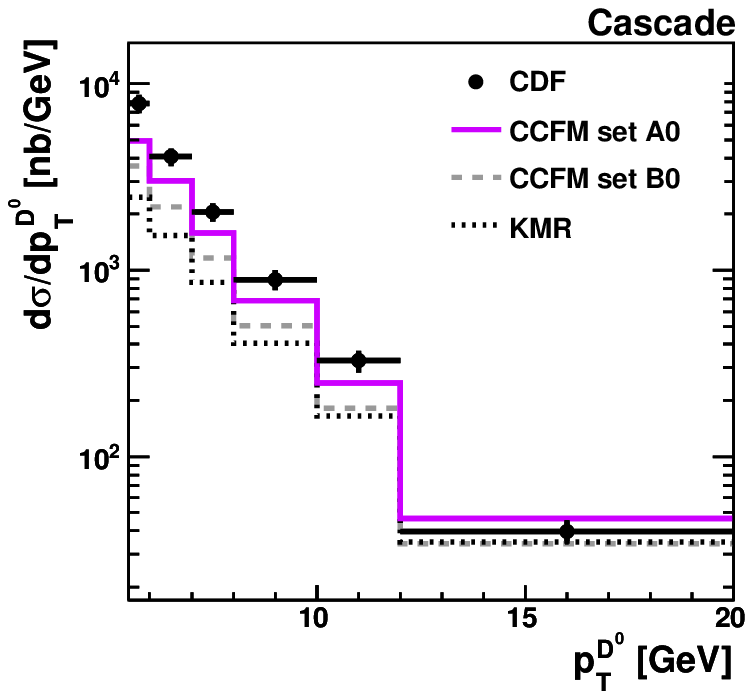, width=10.4cm}}
\put(-2.3,0.5){\epsfig{figure=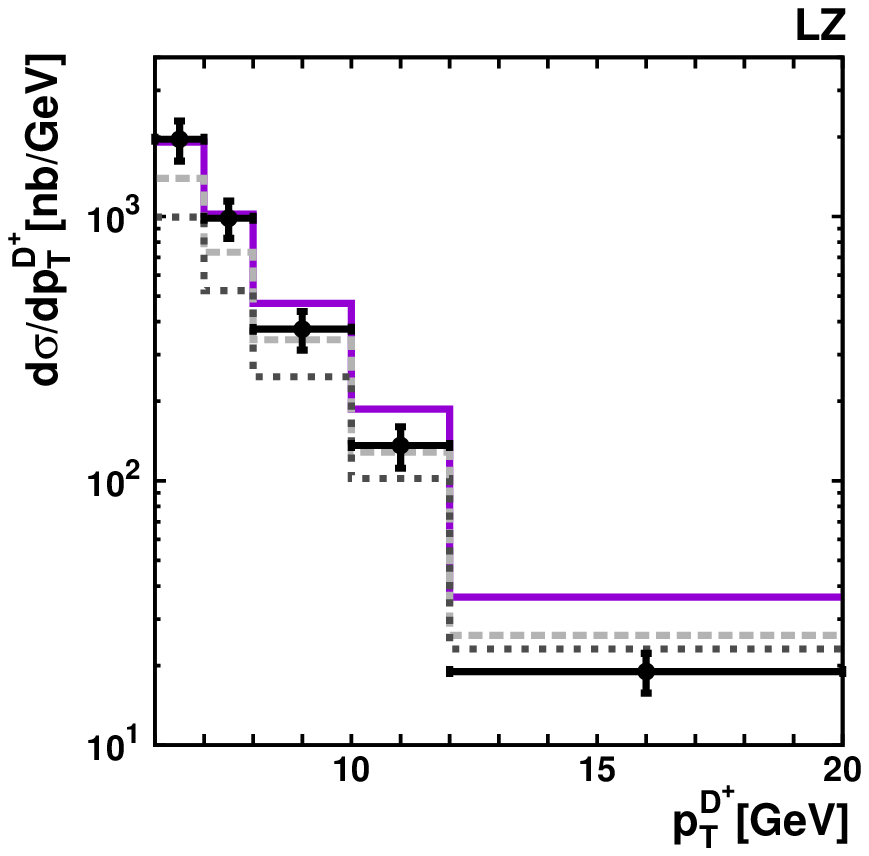, height = 7.8cm, width = 12.2cm}}
\put(8.0,0.0){\epsfig{figure=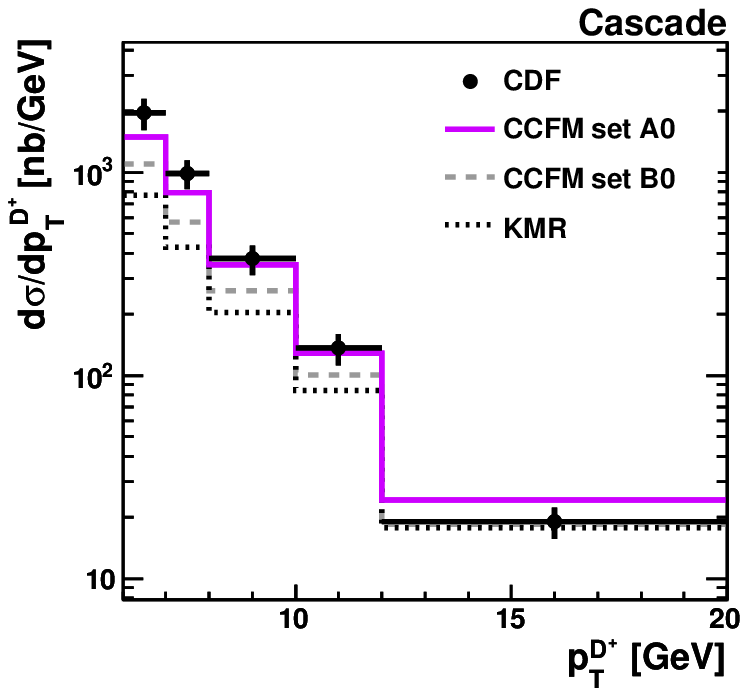, width = 10.4cm}}
\put(4.3,16.){(a)}
\put(12.8,16.){(b)}
\put(4.3,8.){(c)}
\put(12.8,8.){(d)}
\end{picture}
\caption{The cross section as a function of transverse momentum of 
$D^{0}$ [(a), (b)] and $D^{+}$ [(c), (d)] meson hadroproduction.
The kinematical cuts applied are described in the text. 
The first column shows the LZ numerical results while the second one 
depicts the \textsc{Cascade} predictions.
Notation of all histograms is the same as in Fig.~\protect\ref{fig3}.
The experimental data are from CDF~\protect\cite{5}.}
\label{fig5}
\end{figure}

\begin{figure}
\centering
\begin{picture}(16.5,15.)(0.,0.)
\put(-2.3,8.55){\epsfig{figure=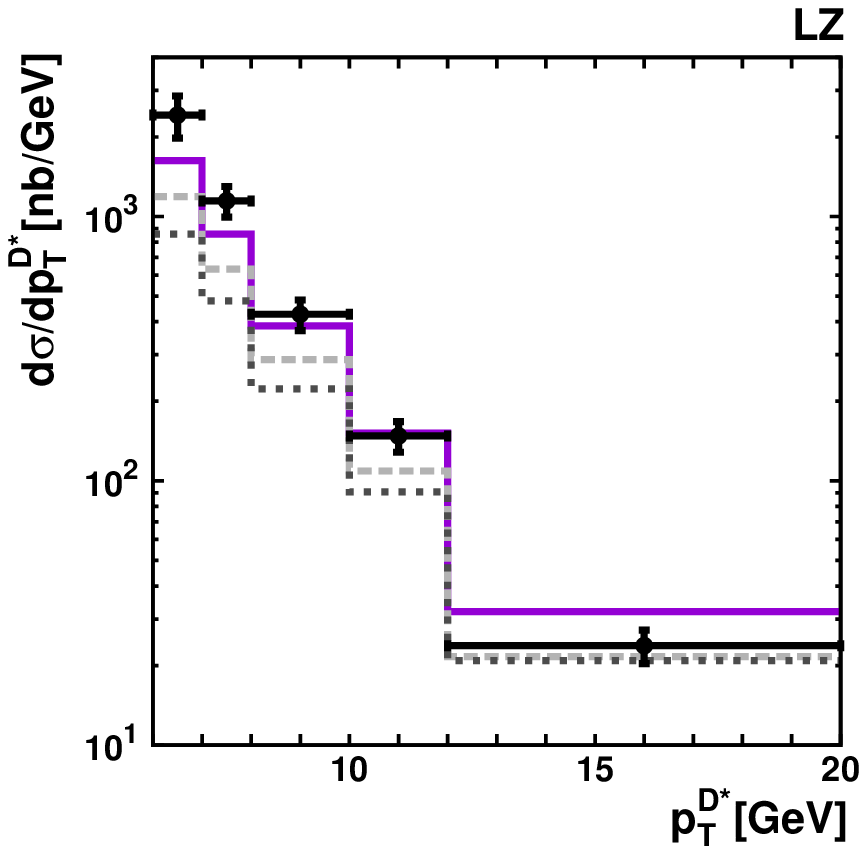, height = 7.8cm, width = 12.2cm}}
\put(8.0,8.){\epsfig{figure=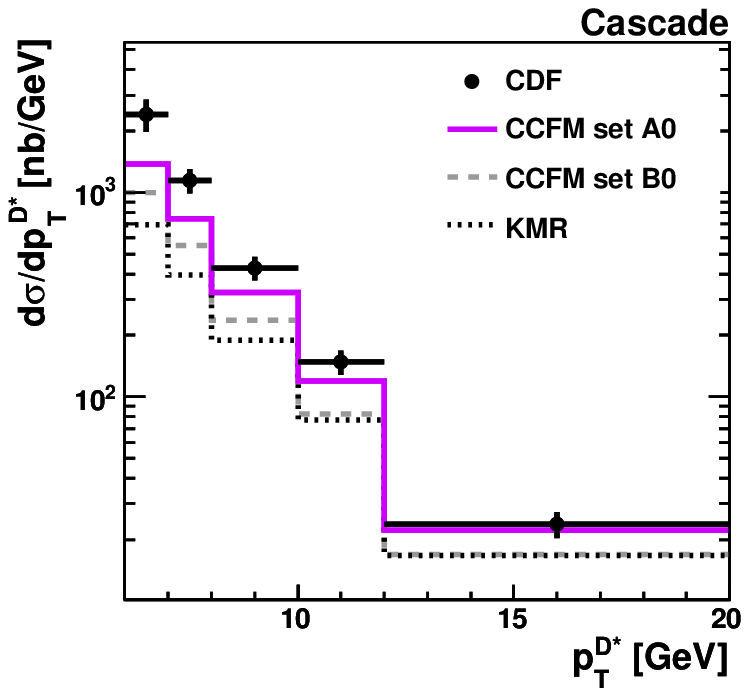, width=10.4cm}}
\put(-2.3,0.5){\epsfig{figure=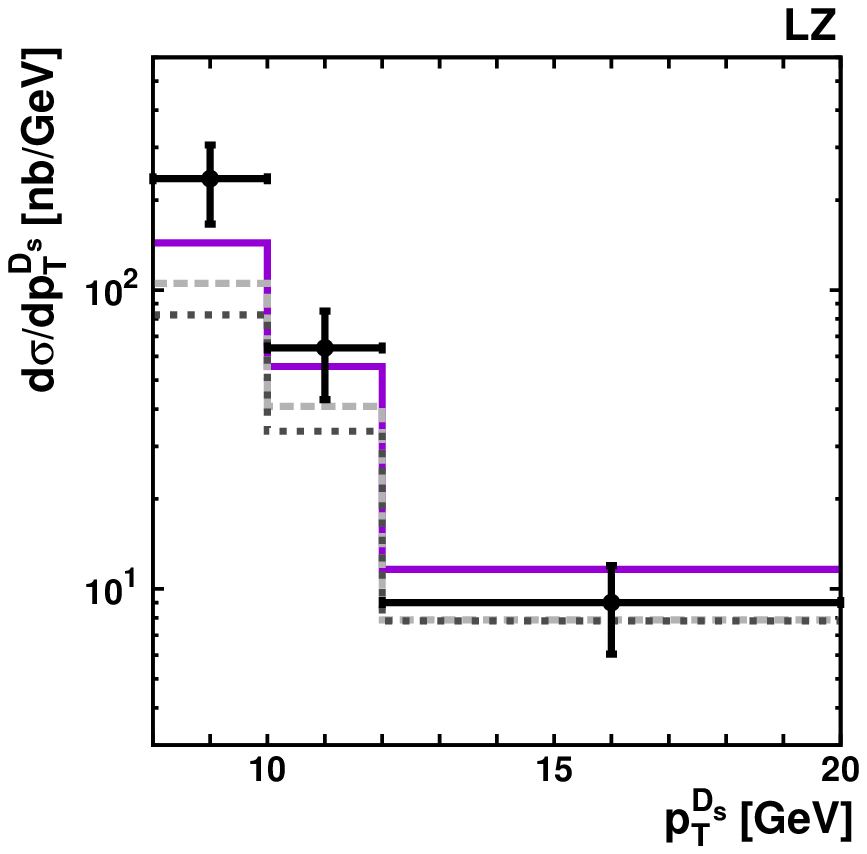, height = 7.8cm, width = 12.2cm}}
\put(8.0,0.0){\epsfig{figure=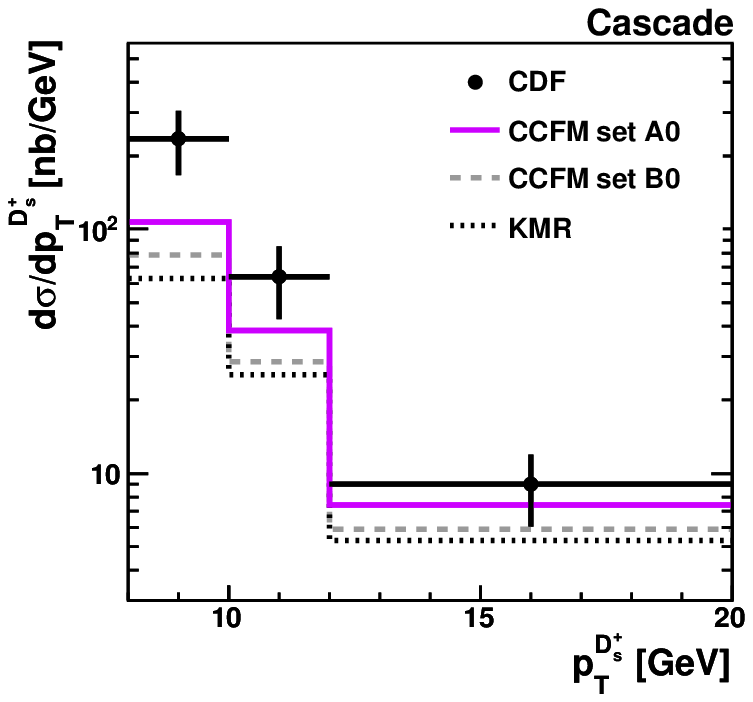, width = 10.4cm}}
\put(4.3,16.){(a)}
\put(12.8,16.){(b)}
\put(4.3,8.){(c)}
\put(12.8,8.){(d)}
\end{picture}
\caption{The cross section as a function of the transverse momentum of 
$D^{*}$ [(a), (b)] and $D^{+}_{s}$ [(c), (d)] meson hadroproduction. 
The kinematical cuts applied are described in the text. 
The first column shows the LZ numerical results while the second one depicts 
the \textsc{Cascade} predictions.
Notation of all histograms is the same as in Fig.~\protect\ref{fig3}.
The experimental data are from CDF~\protect\cite{5}.}
\label{fig6}
\end{figure}

The CDF collaboration has measured~\cite{4,5} the transverse momentum distributions 
of $B^+$ and several $D$ mesons (namely, $D^0$, $D^+$, $D^{*+}$ and $D^+_s$) with $|y| < 1$ (where $y$ is the meson rapidities in
the center-of-mass frame) at $\sqrt s = 1960$~GeV. 
Our predictions are shown in Figs.~\ref{fig4} -- \ref{fig6} in comparison with the data~\cite{4, 5}.
The fragmentation of the charm and beauty quarks 
into a $B$ and $D$ mesons is described with the Peterson fragmentation
function~\cite{31} with  $\epsilon_b = 0.006$
and $\epsilon_c = 0.06$.
According to~\cite{32, 33}, the following branching fractions are used:
$f(b \to B^+)=0.424$,
$f(c \to D^0)=0.582$, $f(c \to D^{+})= 0.268$,
$f(c \to D^{*+})=0.229$ and $f(c \to D^{+}_s)= 0.084$.
The observed difference between the LZ and \textsc{Cascade} predictions
is due to the missing parton shower effects  in the LZ calculations.
We address this point in more detail in Section~\ref{add-effects}.

\begin{figure}
\centering
\begin{picture}(16.5,15.)(0.,0.)
\put(-2.3,8.55){\epsfig{figure=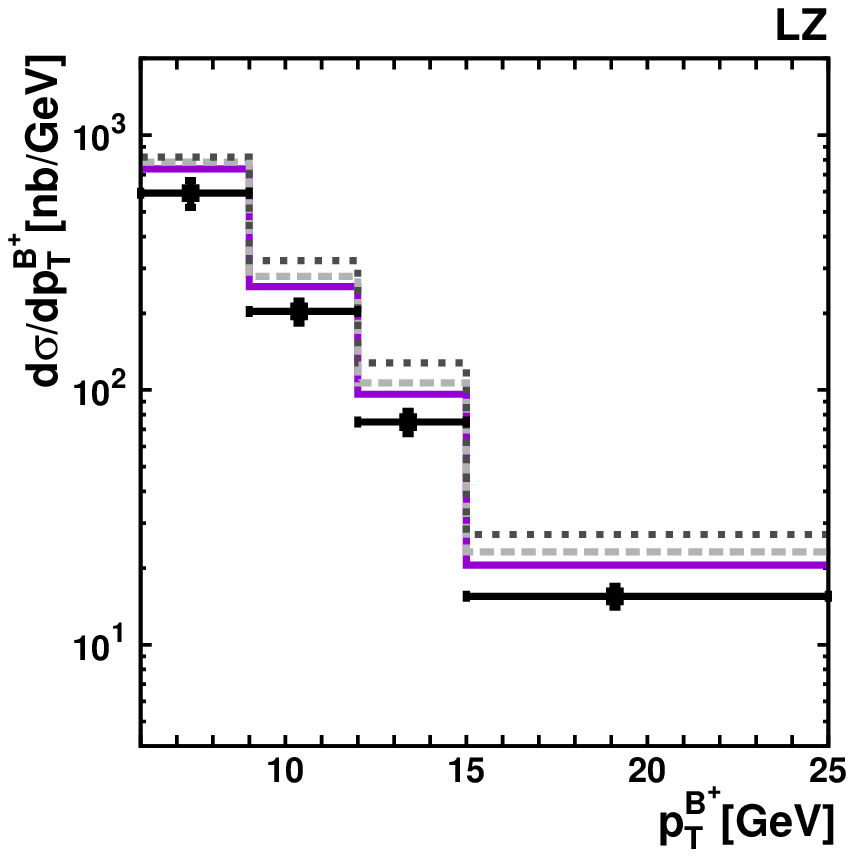, height = 7.8cm, width = 12.2cm}}
\put(8.0,8.0){\epsfig{figure=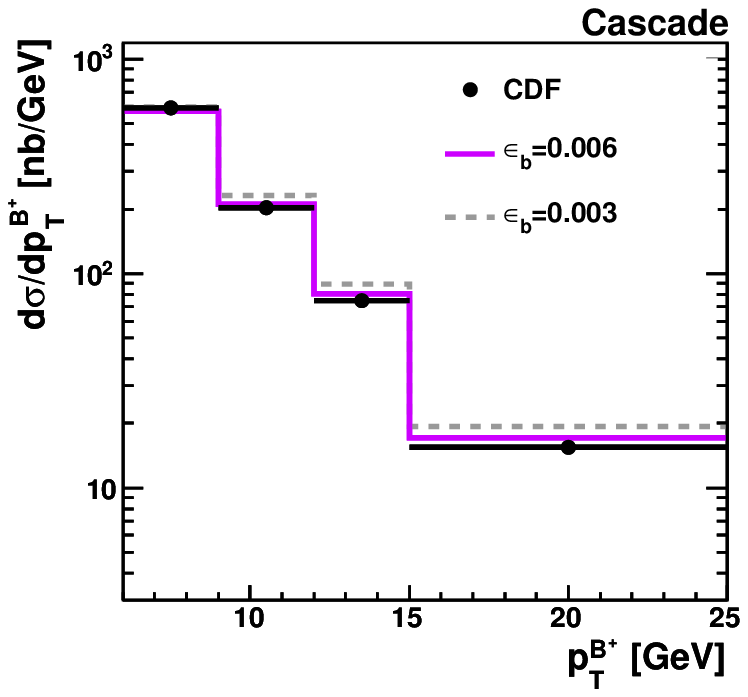, width=10.4cm}}
\put(-2.3,0.5){\epsfig{figure=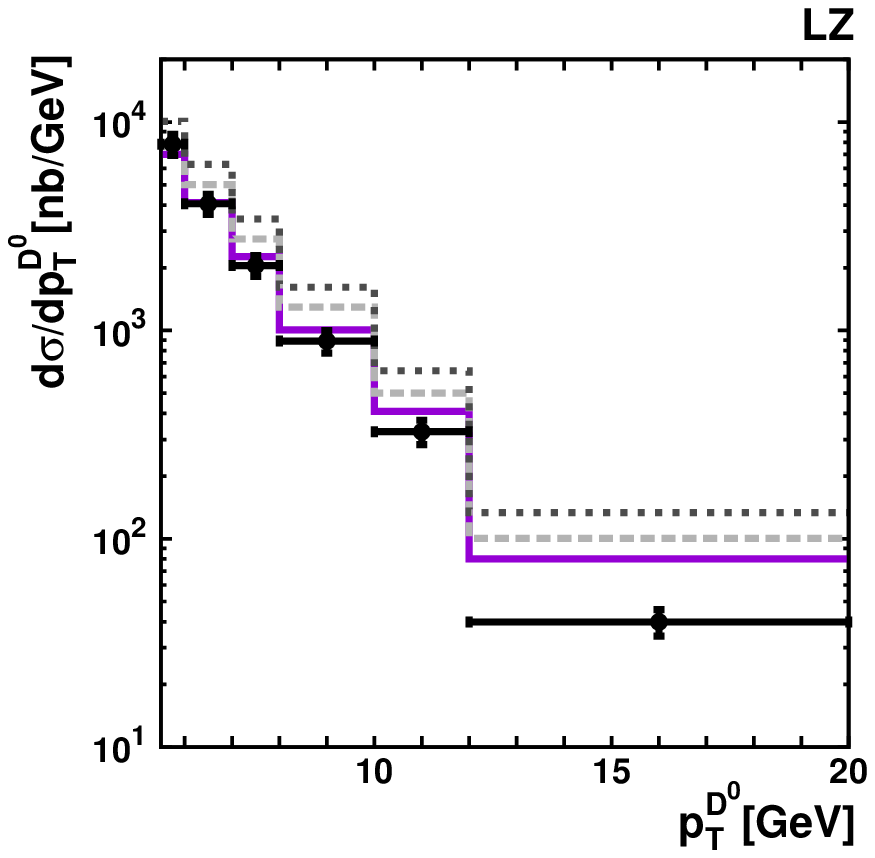, height = 7.8cm, width = 12.2cm}}
\put(8.0,0.0){\epsfig{figure=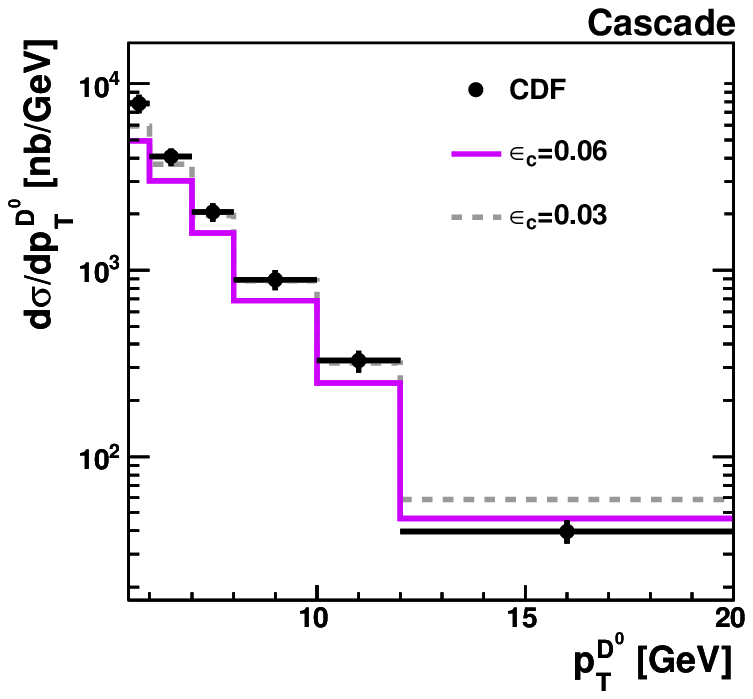, width=10.4cm}}
\put(4.3,16.){(a)}
\put(12.8,16.){(b)}
\put(4.3,8.){(c)}
\put(12.8,8.){(d)}
\end{picture}
\caption{The dependence of our predictions on the fragmentation scheme applied.
The soild, dashed and dotted histograms correspond to the results
obtained using the Peterson fragmentation function with $\epsilon_b = 0.006$
($\epsilon_c = 0.06$), $\epsilon_b = 0.003$ ($\epsilon_c = 0.03$) and
the non-perturbative fragmentation functions from~\protect\cite{7, 8, 34}. 
The first column shows the LZ numerical results while the second one depicts 
the \textsc{Cascade} predictions. Here we
use CCFM set A0 gluon density for illustration.
The experimental data are from CDF~\protect\cite{4, 5}.}
\label{fig7}
\end{figure}

Since the predicted meson transverse momentum distributions
depend on the quark-to-hadron fragmentation, we have repeated our calculations for $B^+$ and $D^0$ mesons 
with the shifted values of the Peterson shape parameter $\epsilon$, namely $\epsilon_b = 0.003$
and $\epsilon_c = 0.03$. These values are also often used in the NLO pQCD calculations.
Additionally, we have applied the non-perturbative fragmentation functions which
have been proposed in~\cite{7, 8, 34} and which have been
used in the FONLL calculations. The input parameters 
were determined~\cite{8, 34} by a  fit to LEP data. The results of our calculations
are shown in Fig.~\ref{fig7}. We find that the predicted cross sections (in the considered $p_T$ region)
are larger for smaller values of parameter $\epsilon$  or
if the fragmentation function from~\cite{7, 8, 34} is used.
However, the typical dependence of numerical
predictions on the fragmentation scheme is
much smaller than the dependence on the 
unintegrated gluon density. 

\begin{figure}
\centering
\begin{picture}(16.5,15.)(0.,0.)
\put(-2.3,8.55){\epsfig{figure=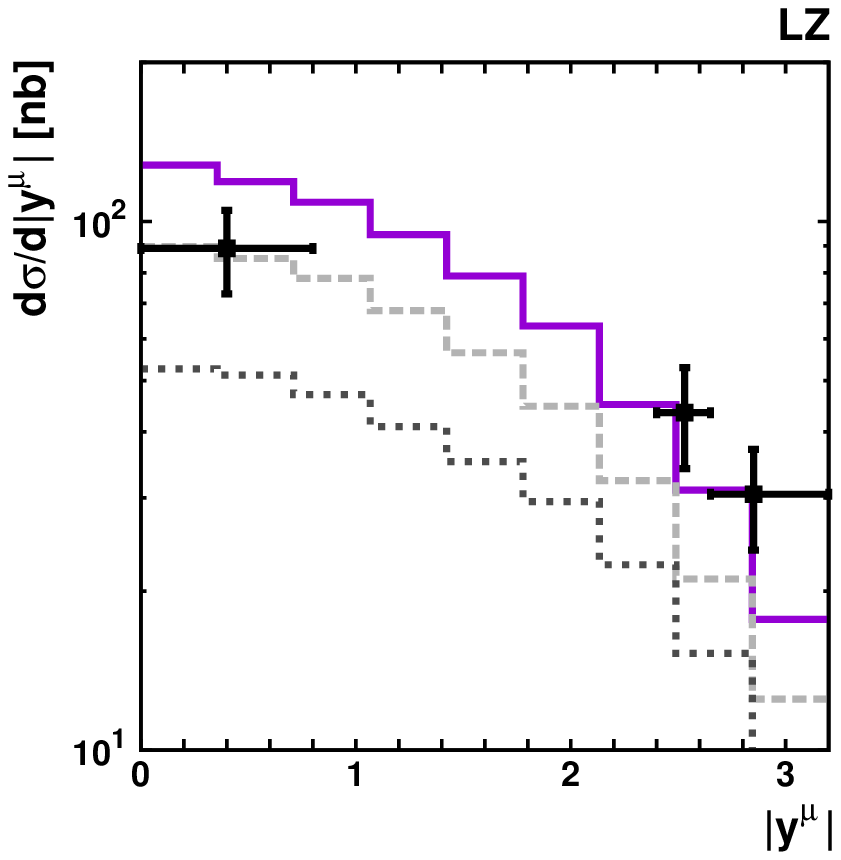, height = 7.8cm, width = 12.2cm}}
\put(7.9,8.0){\epsfig{figure=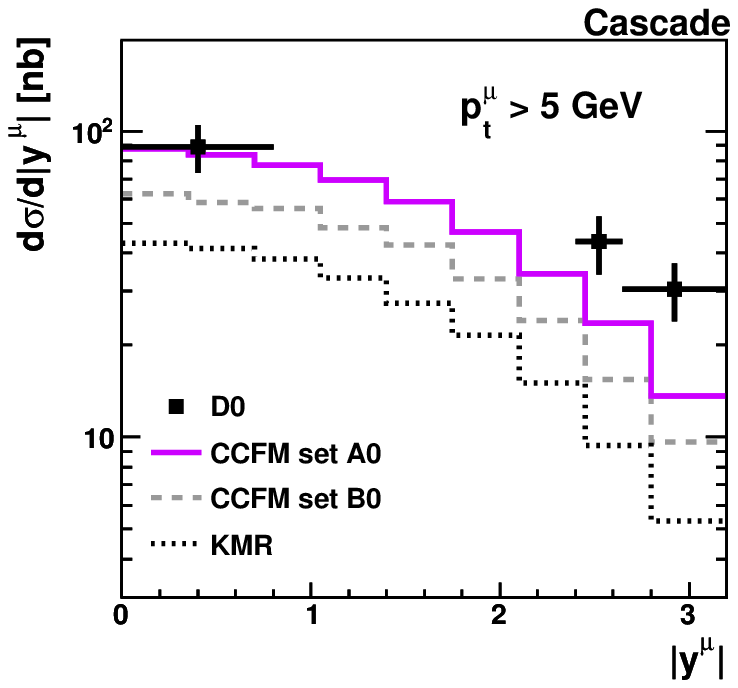, width=10.4cm}}
\put(-2.3,0.5){\epsfig{figure=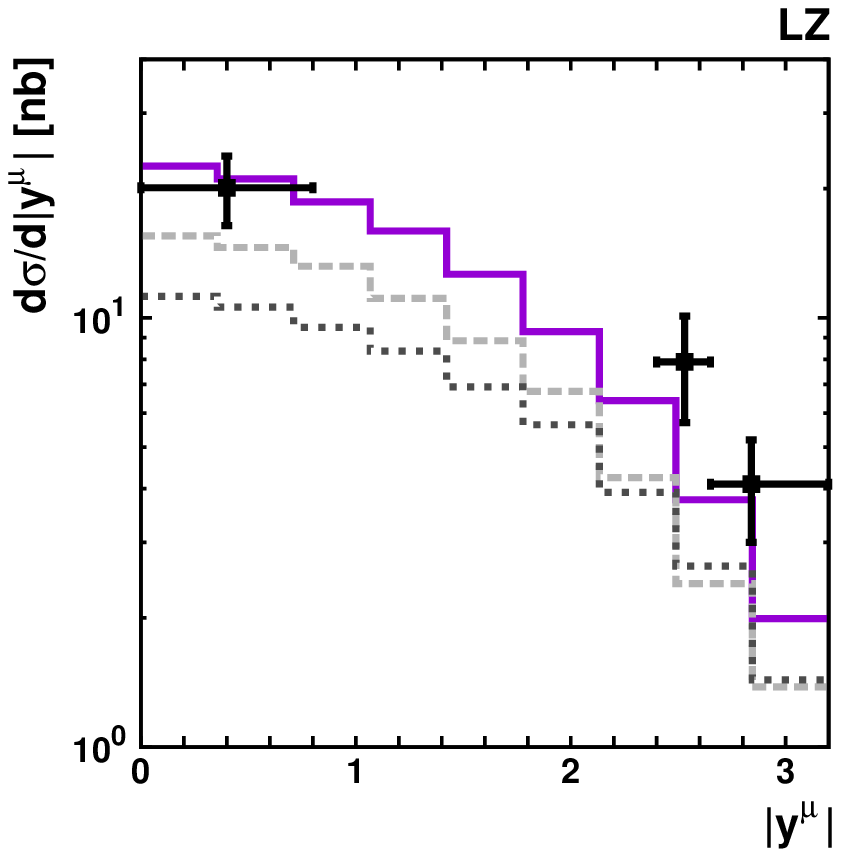, height = 7.8cm, width = 12.2cm}}
\put(7.9,0.0){\epsfig{figure=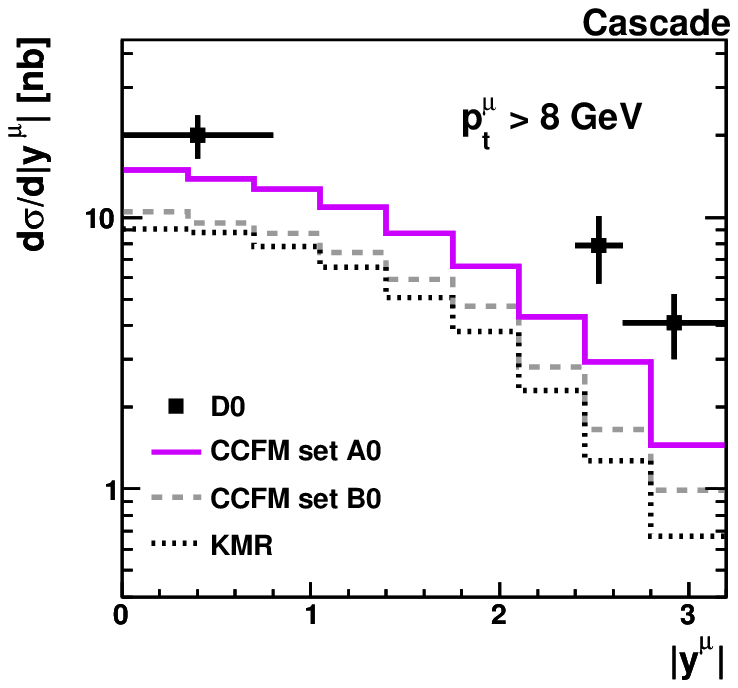, width=10.4cm}}
\put(4.3,16.){(a)}
\put(12.8,16.){(b)}
\put(4.3,8.){(c)}
\put(12.8,8.){(d)}
\end{picture}
\caption{The rapidity distributions of muons arising from the 
semileptonic decays of $B$-mesons. The first column shows the LZ numerical 
results while the second one depicts the \textsc{Cascade} predictions.
The kinematical cuts applied are described in the text.
Notation of all histograms is the same as in Fig.~\protect\ref{fig3}.
The experimental data are from D0~\protect\cite{2}.}
\label{fig8}
\end{figure}

\begin{figure}
\centering
\begin{picture}(16.5,8.)(0.,0.)
\put(-2.3,0.6){\epsfig{figure=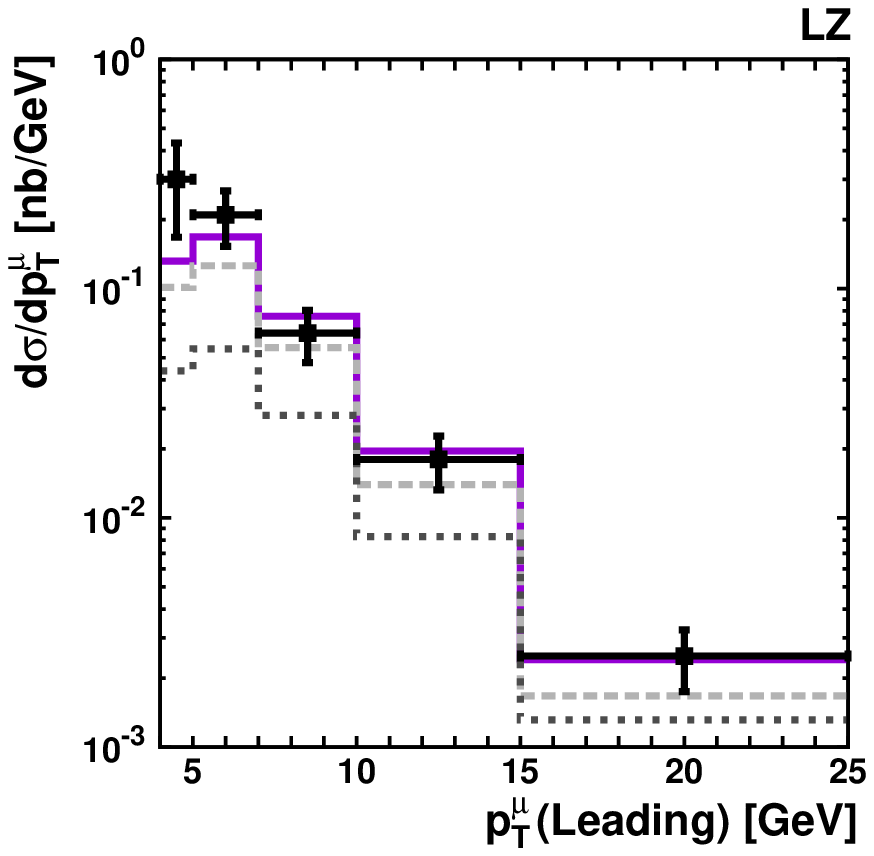, height = 7.7cm, width = 12.2cm}}
\put(8.0,0.0){\epsfig{figure=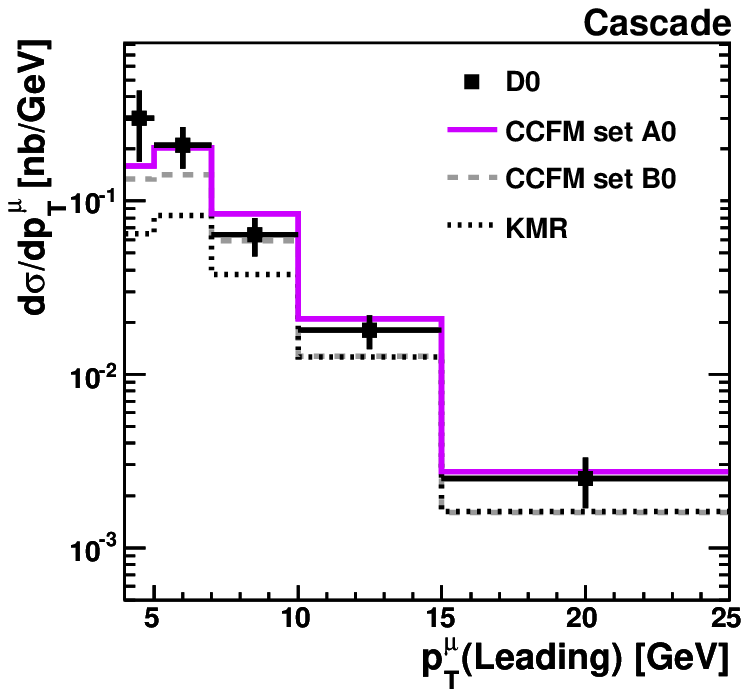, width = 10.4cm}}
\put(4.3,8.){(a)}
\put(12.8,8.){(b)}
\end{picture}
\caption{The transverse momentum distributions of muons arising from the 
semileptonic decays of $B$-mesons. The left histogram shows the LZ numerical 
results while the right plot depicts the \textsc{Cascade} predictions.
The kinematical cuts applied are described in the text.
Notation of all histograms is the same as in Fig.~\protect\ref{fig3}.
The experimental data are from D0~\protect\cite{1}.}
\label{fig9}
\end{figure}

The D0 experiment has measured muons originating
from the semileptonic decays of $b$-quarks at $\sqrt s = 1800$~GeV~\cite{1}
 for  $4 < p_T^\mu < 25$ GeV, 
$|\eta^\mu| < 0.8$ (for both muons) and $6 < m^{\mu \mu} < 35$~GeV, 
where $\eta^\mu$ is the muon pseudo-rapidity and 
$m^{\mu \mu}$ is the invariant mass of the produced muon pair.
In~\cite{2} the measurements are extended to the forward muon rapidity region, namely $2.4 < |y^\mu| < 3.2$.
To produce muons from $b$-quarks 
in the LZ calculations, we first convert $b$-quarks into $B$-mesons 
(using the Peterson fragmentation function with default value $\epsilon_b = 0.006$) 
and then simulate their semileptonic 
decay according to the standard electroweak theory.
The branching of $b \to  \mu $ as well as the cascade decay $b\to c\to \mu$ are taken into account with the branching fraction taken from~\cite{33}.
The predictions of the LZ and {\sc Cascade} calculations are shown in 
Figs.~\ref{fig8} and ~\ref{fig9}.
We find that our predictions with both 
CCFM-evolved unintegrated gluon densities describe the experimental
data for both the transverse momentum and rapidity distributions of 
muons reasonably well.

The calculated total cross sections of the $b$-quarks, $B$ and $D$ mesons
and their decay muons compared to the CDF experimental data~\cite{4, 5} are 
listed in Table \ref{table_pdfs}. In Table~\ref{table_uncertainties}
the systematic uncertainties of our calculations are summarized.
To estimate the uncertainty coming from the renormalization scale $\mu_R$, we used 
the CCFM set A0$+$ and A0$-$ instead of the default density function A0 in {\sc Cascade}. 
These two sets represent a variation of the scale used in $\alpha_s$ in the 
off-shell matrix element.
The A0$+$ stands
for a variation of $2\,\mu_R$, while set A0$-$ reflects $\mu_R/2$. 
In all heavy flavour analyses studied 
here, we observe a deviation of roughly $+10\%$ for set A0$+$. The uncertainty coming
from set A0$-$ is generally smaller, but still positive.
The dependence on the heavy flavour masses is investigated as well using {\sc Cascade}. 
We varied our default values of $m_b=4.75$~GeV by  $\pm0.25$ GeV 
and $m_c=1.4$~GeV by $\pm0.1$~GeV. All heavy flavour cross 
sections are observed to be lower, if the masses are enlarged and vice versa.  

\begin{table}
\begin{center}
\begin{tabular}{|l|c|c|c|c|c|}
\hline
   & & & & & \\
  Source & $\sigma(B^{+})$ & $\sigma(D^{0})$ &  $\sigma(D^{+})$ & $\sigma(D^{*+})$ & $\sigma(D^{+}_s)$ \\
   & & & & &  \\
\hline
   & & & & & \\
   CDF data [$\mu$b]  & $2.78 \pm 0.24$ & $13.3 \pm 1.5$ & $4.3 \pm 0.7$ & $5.2 \pm 0.8$  & $0.75 \pm 0.23$ \\
   & & & & & \\
   \hline
   & & & & & \\
   A0 (LZ/\textsc{Cascade}) & 3.47/2.76 & 13.34/9.31 & 4.53/3.45 & 3.81/3.19 & 0.49/0.35 \\
   & & & & & \\
   B0 (LZ/\textsc{Cascade}) & 2.54/2.02 & 9.55/6.78 & 3.27/2.54 & 2.78/2.32 & 0.35/0.26 \\
   & & & & & \\
   KMR (LZ/\textsc{Cascade}) & 1.33/1.16 & 6.92/5.06 & 2.40/1.92 & 2.13/1.75 & 0.29/0.22 \\
   & & & & & \\
   KMR (LZ, MSTW 2008) & 1.16 & 6.04 & 2.17 & 1.79 & 0.25 \\
   & & & & & \\
   \hline
\end{tabular}
\end{center}
\caption{The charm and beauty total cross section in 
  $p\bar p$ collisions at $\sqrt s = 1960$~GeV. The measurement~\protect\cite{4, 5} was obtained
  in the central rapidity region ($|y| < 1$) at $6 < p_T < 25$~GeV and
  $5.5 < p_T < 20$~GeV in the case of $B$ and $D$ mesons, respectively.}\label{table_pdfs}
\end{table}
   
   \begin{table}
\begin{center}
\begin{tabular}{|l|c|c|c|c|c|}
\hline
   & & & & & \\
  Source & $\sigma(B^{+})$ & $\sigma(D^{0})$ &  $\sigma(D^{+})$ & $\sigma(D^{*+})$ & $\sigma(D^{+}_s)$ \\
   & & & & &  \\
\hline
   & & & & & \\
   CCFM set A0& 2.76 & 9.31 & 3.45 & 3.19 & 0.35 \\
   & & & & & \\
   CCFM set A0$+$ & +10\% & +6\%  & +6\% & +5\% & +9\% \\
   & & & & & \\
   CCFM set A0$-$ & +1\% & +4\% & +4\% & +3\% & +3\% \\
   & & & & & \\
   $m_b=5.0$ GeV, $m_c=1.5$ GeV  & -8\% & -4\% & -4\% & -4\% & -3\% \\
   & & & & & \\
    $m_b=4.5$ GeV, $m_c=1.3$ GeV & +11\% & +4\%  & +1\% & +4\% & +11\% \\
   & & & & & \\
   $\epsilon_b=0.003$, $\epsilon_c=0.03$ & +7\% & +21\% & +23\% & +21\% & +25\% \\
   & & & & & \\
   \hline
   & & & & & \\
   Total & $\pm^{16\%}_{8\%}$ & $\pm^{23\%}_{4\%}$ &$\pm^{24\%}_{4\%}$ &  $\pm^{22\%}_{4\%}$& $\pm^{29\%}_{3\%}$ \\
   & & & & & \\
\hline
\end{tabular}
\end{center}
\caption{Systematic uncertainties for charm and beauty total cross sections in 
  $p\bar p$ collisions at $\sqrt s = 1960$~GeV obtained with  \textsc{Cascade}. The measurement~\protect\cite{4, 5} was obtained
  in the central rapidity region ($|y| < 1$) at $6 < p_T < 25$~GeV and
  $5.5 < p_T < 20$~GeV in the case of $B$ and $D$ mesons, respectively.}\label{table_uncertainties}
\end{table}

\begin{figure}
\centering
\begin{picture}(16.5,15.)(0.,0.)
\put(-2.3,8.55){\epsfig{figure=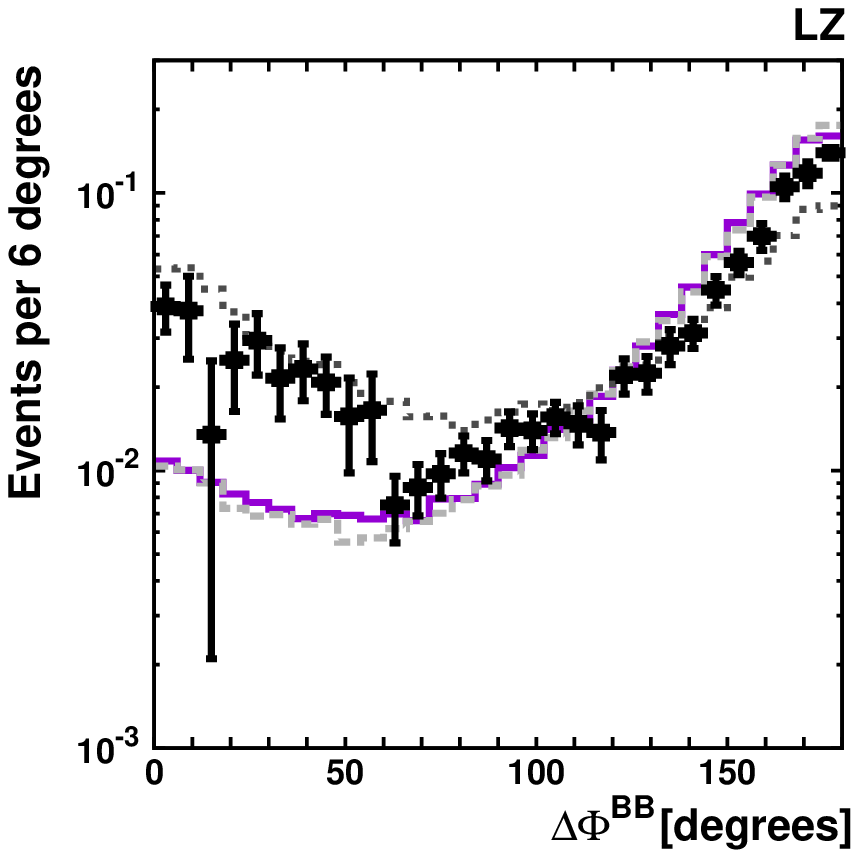, height = 7.8cm, width = 12.2cm}}
\put(8.0,8.0){\epsfig{figure=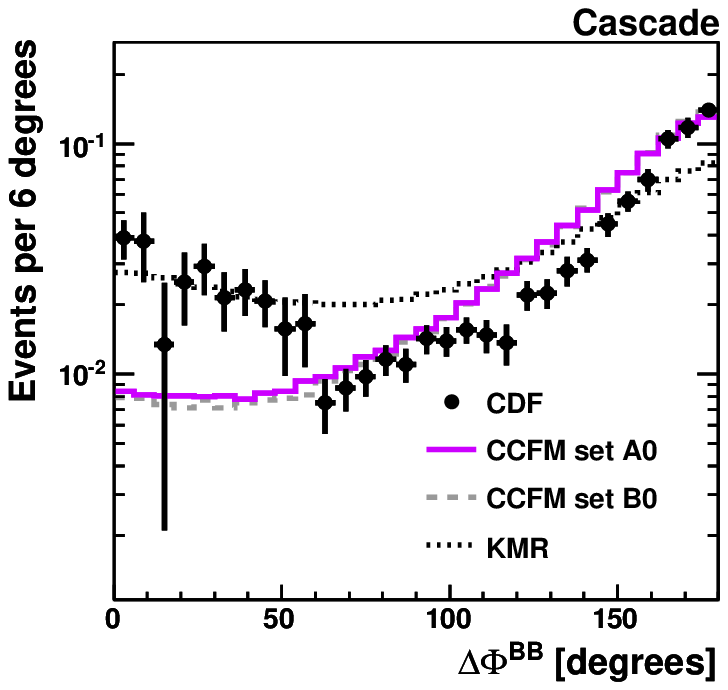, width=10.4cm}}
\put(-2.3,0.5){\epsfig{figure=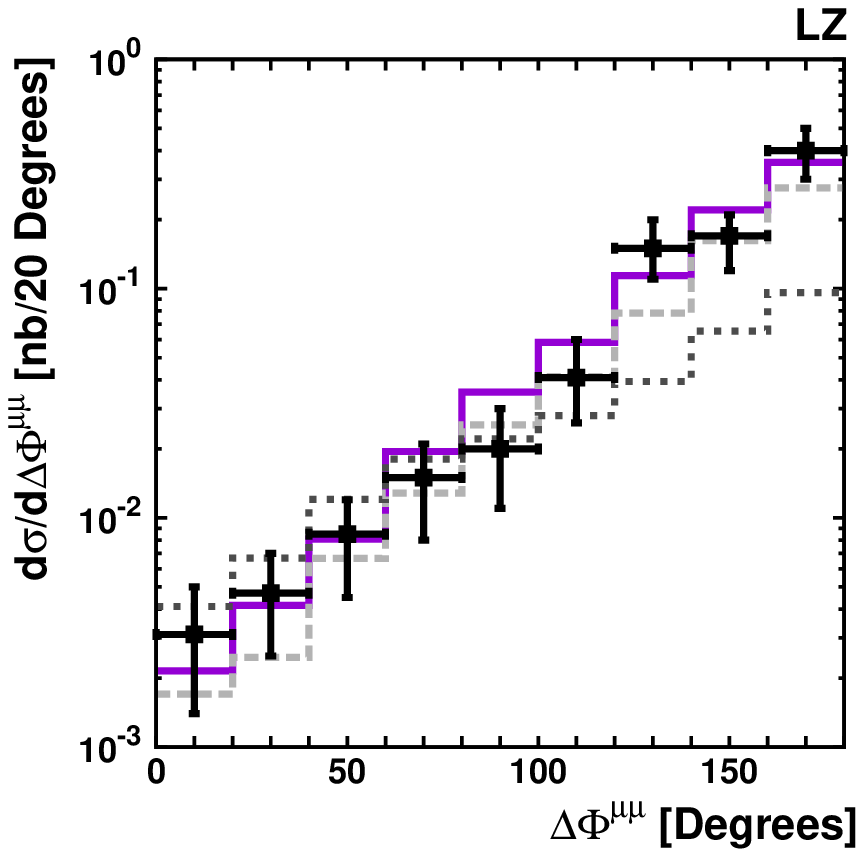, height = 7.8cm, width = 12.2cm}}
\put(8.0,0.0){\epsfig{figure=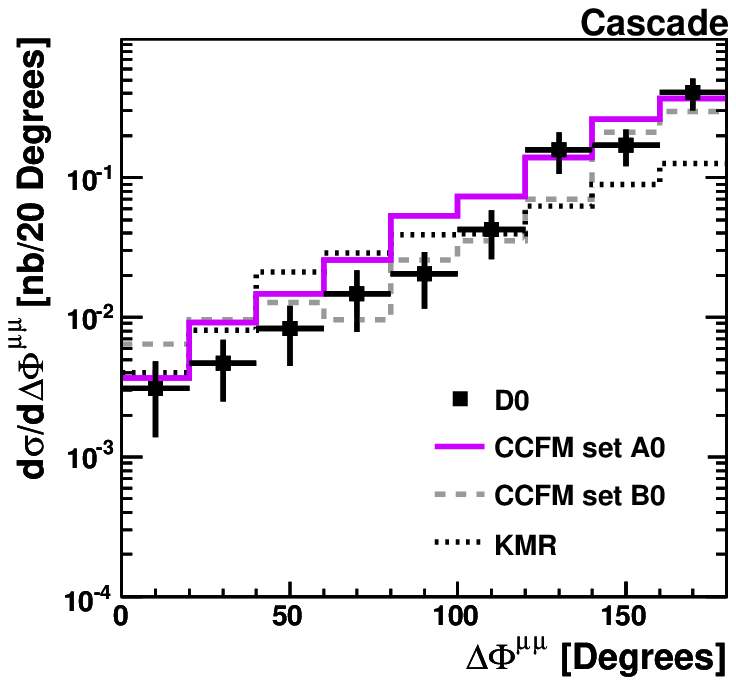, width=10.4cm}}
\put(4.3,16.){(a)}
\put(12.8,16.){(b)}
\put(4.3,8.){(c)}
\put(12.8,8.){(d)}
\end{picture}
\caption{The distributions in azimuthal angle difference
between the produced $B$-mesons [(a), (b)] and muons [(c), (d)], 
originating from their semileptonic decays.
The first column shows the LZ numerical 
results while the second one depicts the \textsc{Cascade} predictions.
The kinematical cuts applied are described in the text.
Notation of all histograms is the same as in Fig.~\protect\ref{fig3}.
The experimental data are from CDF~\protect\cite{3} and D0~\protect\cite{1}.}
\label{fig10}
\end{figure}

\begin{figure}
\centering
\begin{picture}(16.5,15.)(0.,0.)
\put(-2.3,8.55){\epsfig{figure=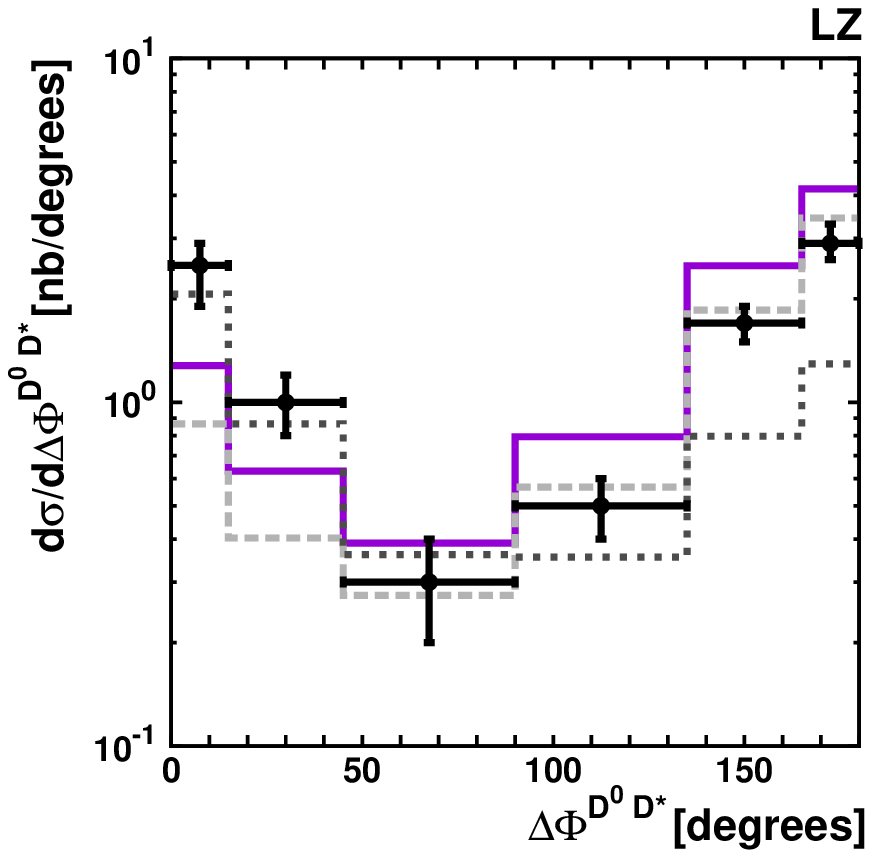, height = 7.8cm, width = 12.2cm}}
\put(7.9,8.){\epsfig{figure=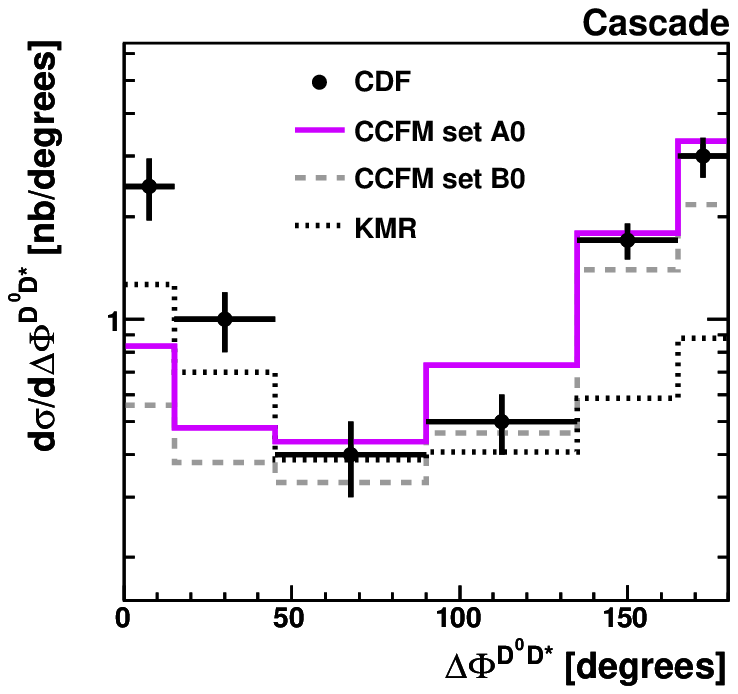, width=10.4cm}}
\put(-2.3,0.5){\epsfig{figure=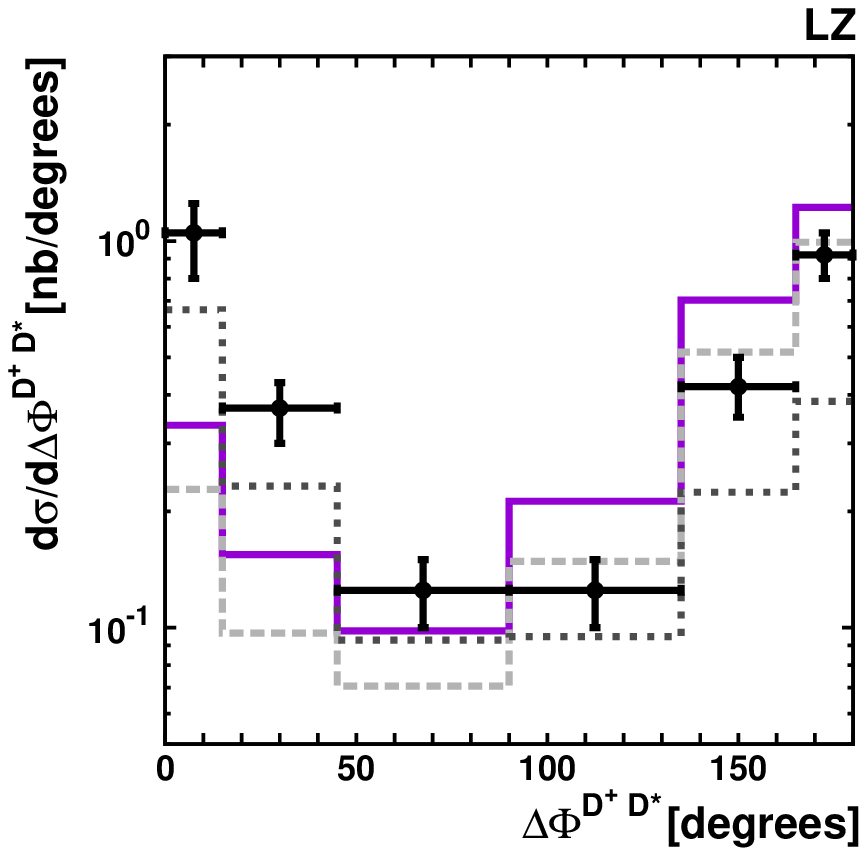, height = 7.8cm, width = 12.2cm}}
\put(7.9,0.0){\epsfig{figure=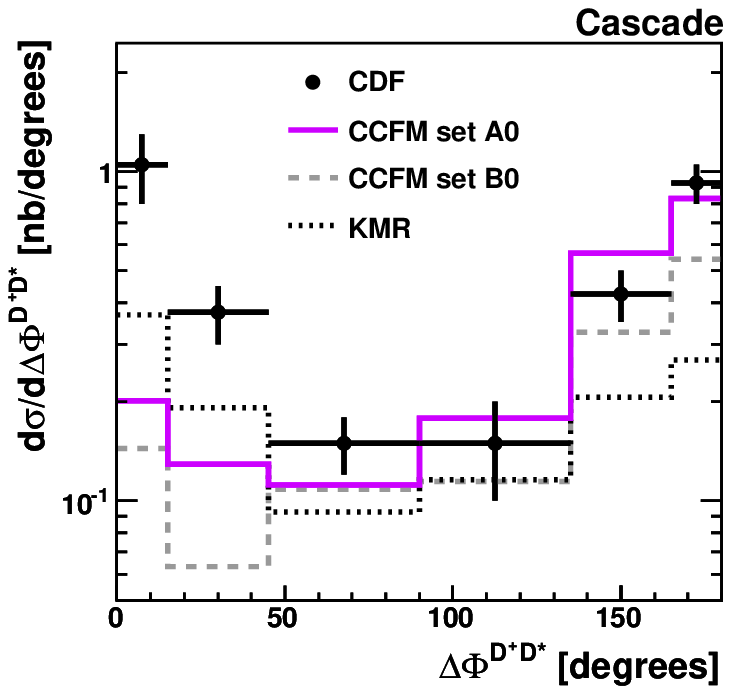, width=10.4cm}}
\put(4.3,16.){(a)}
\put(12.8,16.){(b)}
\put(4.3,8.){(c)}
\put(12.8,8.){(d)}
\end{picture}
\caption{The distributions in azimuthal angle difference
between the produced $D^0$, $D^{*-}$ and 
$D^+$, $D^{*-}$ mesons. The first column shows the LZ numerical 
results while the second one depicts the \textsc{Cascade} predictions.
The cuts applied are described in the text.
Notation of all histograms is the same as in Fig.~\protect\ref{fig3}.
The experimental data are from CDF~\protect\cite{6}.}
\label{fig11}
\end{figure}

We turn now to the investigation of
the angular correlations between the produced
particles. As it was mentioned above, such correlations have been not
studied yet in the framework of GM-VFNS scheme (which has been 
applied for transverse momentum distributions of the charm
and beauty mesons only~\cite{9, 10}). Experimental data on the azimuthal 
correlations in charm and beauty
production come from both the CDF and D0 collaborations.
In the case of $b$-quark production, CDF data~\cite{3} refer to
the $B, \bar B$ azimuthal angle distribution measured at 
$|y| < 1$, $p_T(B) > 14$~GeV, 
$p_T(\bar B) > 7.5$~GeV and $\sqrt s = 1800$~GeV. The D0
data~\cite{1} refer to the muon-muon correlation measured in the region of
$4 < p_T^\mu < 25$~GeV, $|\eta^\mu| < 0.8$ and 
$6 < m^{\mu \mu} < 35$~GeV at the same energy.
In the case of charm production, the
CDF collaboration has presented the measurement~\cite{6} of 
$D^0$, $D^{*-}$ and  
$D^+$, $D^{*-}$ pair cross section as a function of angle
between the two charm mesons for the kinematic range
$|y| < 1$, $5.5 < p_T(D^0) < 20$~GeV, $5.5 < p_T(D^{*-}) < 20$~GeV
and $7 < p_T(D^{+}) < 20$~GeV. 
Our predictions are shown in Figs.~\ref{fig10} and \ref{fig11} in comparison with 
the data~\cite{1, 3, 6}. 
We observe that
the predicted shapes of azimuthal angle distributions are very 
different for different unintegrated gluon density functions.
This is in a contrast to the cross sections as a function of transverse
momenta or rapidities where all unintegrated gluon densities  gave a  similar behaviour.
We can conclude that the cross section as a
function of $\Delta \phi$ is very sensitive to the details
of the non-collinear gluon evolution in a proton.
As it was pointed out in~\cite{11, 15, 16}, such observables can serve as an 
important and crucial test discriminating the different approaches
of the small-$x$ physics.
The CCFM-evolved gluon densities overshoot the data 
at $\Delta \phi \sim \pi$ and tends to underestimate them at $\Delta \phi \sim 0$.
We observe that the peak at $\Delta \phi \to 0$ is not at all 
reproduced by the CCFM unintegrated gluon densities; we address 
this point in more detail in Section~\ref{add-effects} where we 
discuss the the process $gg^* \to gg$. 
However the KMR gluon density has obviously a very 
different $k_T$ distribution (see Fig.~\ref{fig1}) and therefore provides a better 
description.

\subsection{$b\bar b$ di-jet production} \indent

\begin{figure}
\centering
\begin{picture}(16.5,8.)(0.,0.)
\put(-2.3,0.6){\epsfig{figure=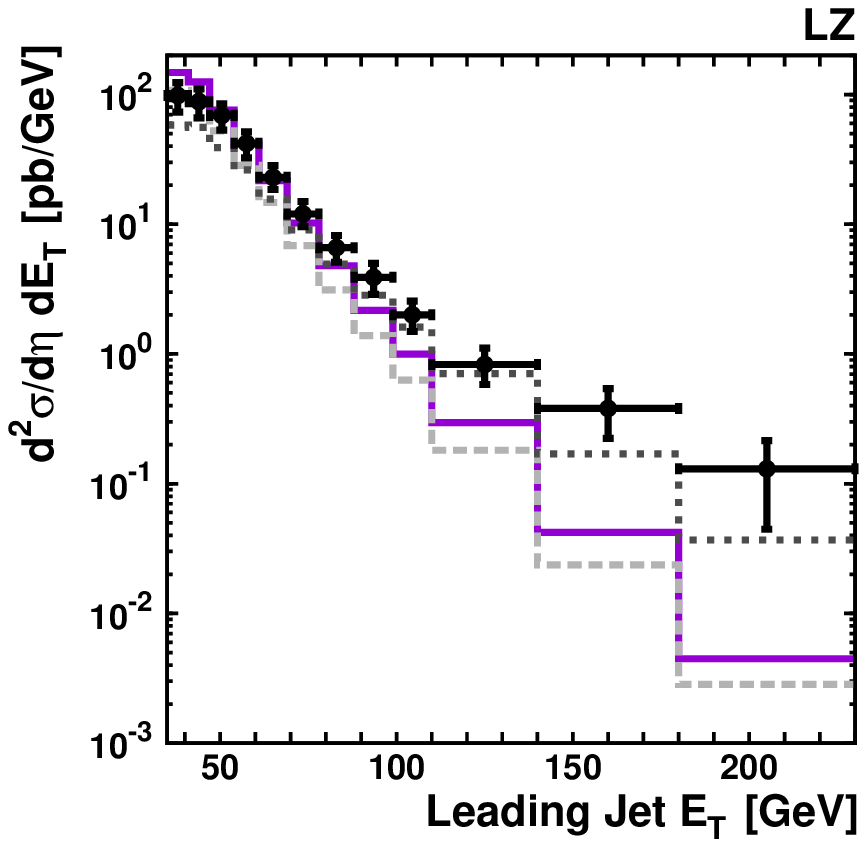, height = 7.7cm, width = 12.2cm}}
\put(7.9,0.0){\epsfig{figure=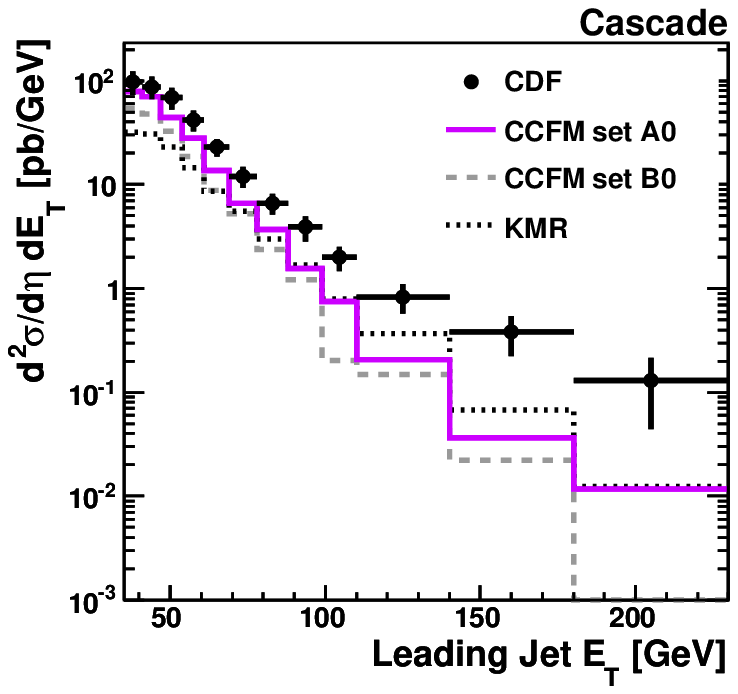, width = 10.4cm}}
\put(4.3,8.){(a)}
\put(12.8,8.){(b)}
\end{picture}
\caption{The leading jet transverse energy distributions of $b\bar b$ di-jet production. 
The left histogram shows the LZ numerical 
results while the right plot depicts the \textsc{Cascade} predictions.
The kinematical cuts applied are described in the text.
Notation of all histograms is the same as in Fig.~\protect\ref{fig3}.
The experimental data are from CDF~\protect\cite{21}.}
\label{fig12}
\end{figure}

\begin{figure}
\centering
\begin{picture}(16.5,8.)(0.,0.)
\put(-2.3,0.6){\epsfig{figure=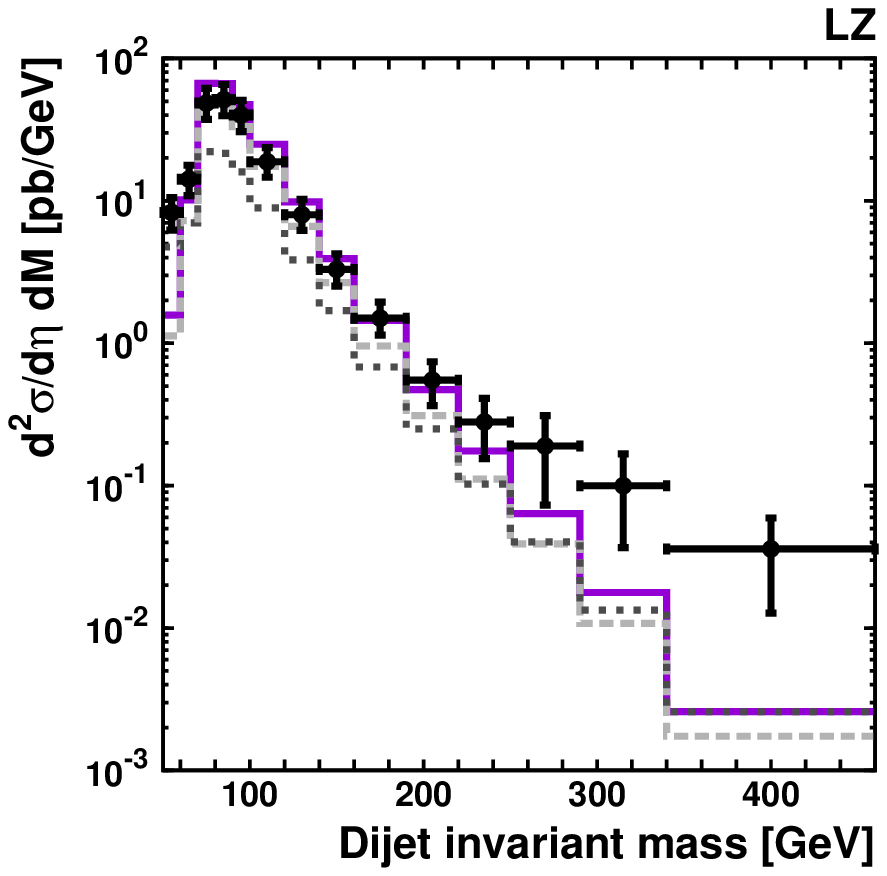, height = 7.7cm, width = 12.2cm}}
\put(8.0,0.0){\epsfig{figure=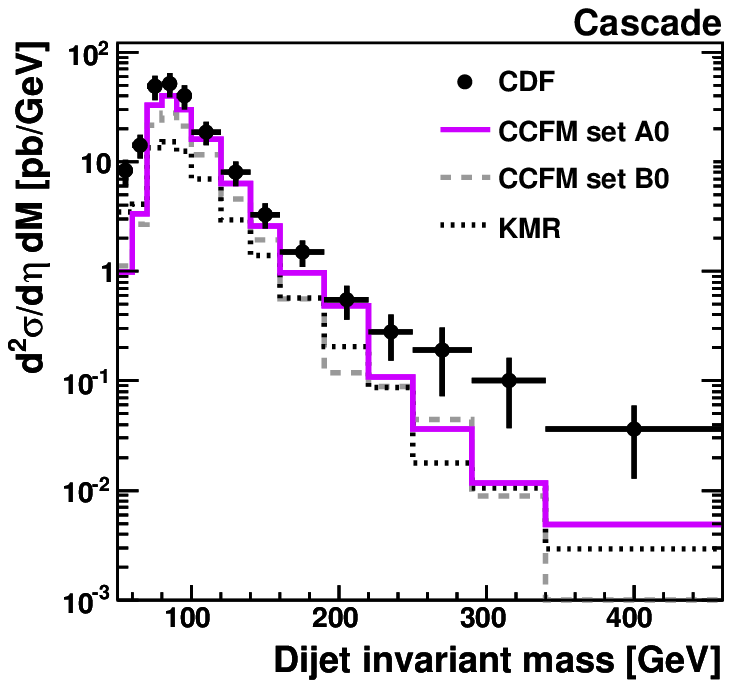, width = 10.4cm}}
\put(4.3,8.){(a)}
\put(12.8,8.){(b)}
\end{picture}
\caption{The invariant mass distributions of $b\bar b$ di-jet production. 
The left histogram shows the LZ numerical 
results while the right plot depicts the \textsc{Cascade} predictions.
The kinematical cuts applied are described in the text.
Notation of all histograms is the same as in Fig.~\protect\ref{fig3}.
The experimental data are from CDF~\protect\cite{21}.}
\label{fig13}
\end{figure}

\begin{figure}
\centering
\begin{picture}(16.5,8.)(0.,0.)
\put(-2.3,0.6){\epsfig{figure=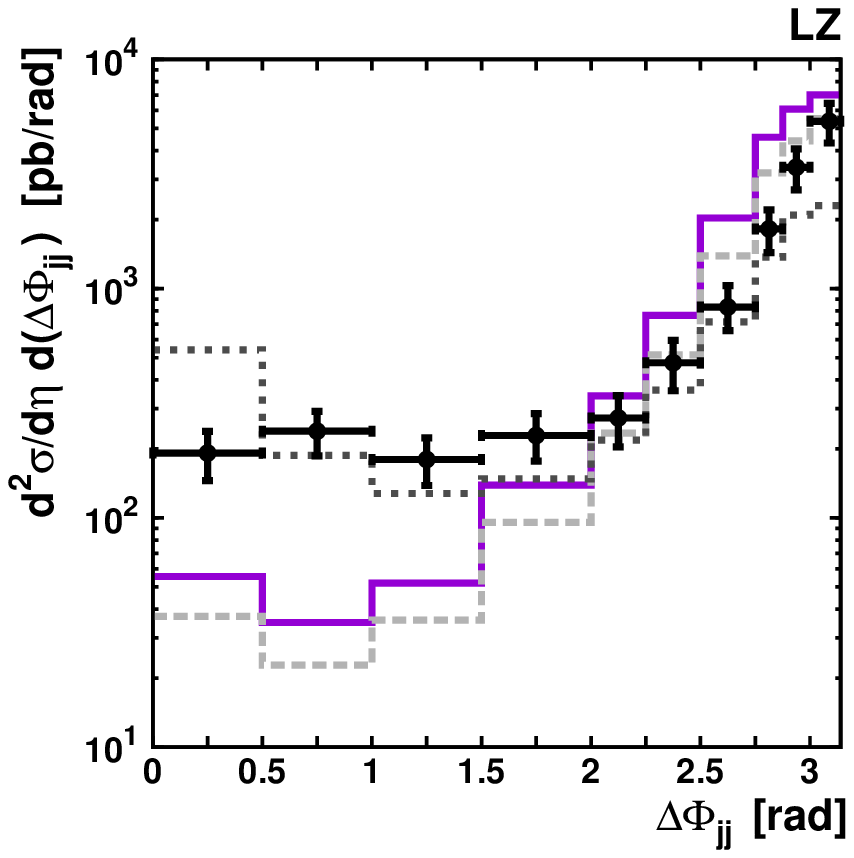, height = 7.7cm, width = 12.2cm}}
\put(7.9,0.0){\epsfig{figure=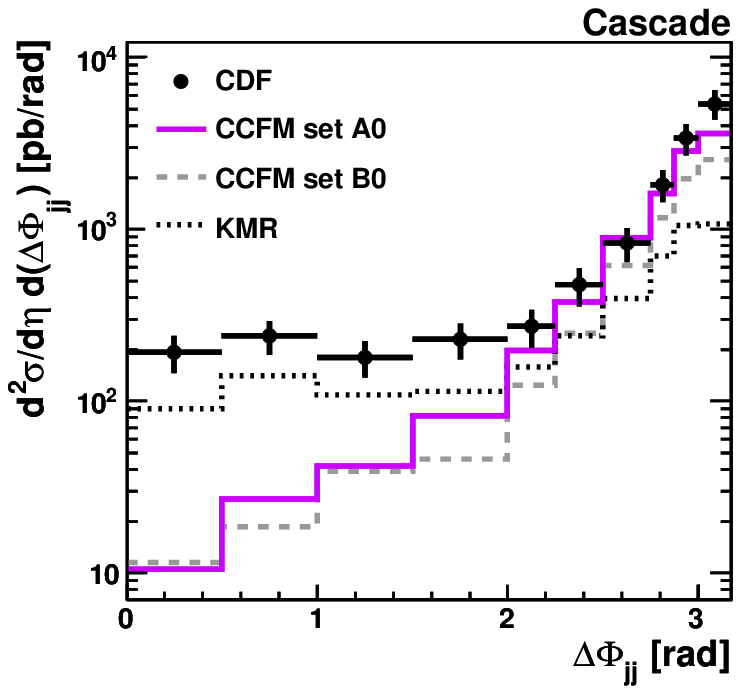, width = 10.4cm}}
\put(4.3,8.){(a)}
\put(12.8,8.){(b)}
\end{picture}
\caption{
The distributions in azimuthal angle difference
between the produced $b$-jets.
The left histogram shows the LZ numerical 
results while the right plot depicts the \textsc{Cascade} predictions.
The kinematical cuts applied are described in the text.
Notation of all histograms is the same as in Fig.~\protect\ref{fig3}.
The experimental data are from CDF~\protect\cite{21}.}
\label{fig14}
\end{figure}

Recently, the experimental study of the $b\bar b$ di-jet production
in $p\bar p$ collisions at $\sqrt s = 1960$~GeV has been presented~\cite{21,vallecorsa-2007}.
The total cross section has been measured in the kinematical region
defined by $|y_1| < 1.2$, $|y_2| < 1.2$, $E_{1T} > 35$~GeV and  $E_{2T} > 32$~GeV.
The differential cross sections as a functions of the leading jet transverse 
energy and of the di-jet invariant mass have been measured with
the azimuthal angle correlation between the two jets. The $b$-jets 
are reconstructed with the JETCLU cone algorithm with radius $R > 0.4$.
Our predictions are shown in Figs.~\ref{fig12} --\ref{fig14}. We observe that at large $E_T$ and at 
large dijet invariant masses $M$ the predictions fall below the measurement. 
Note, however, that in this kinematical region the quark induced 
subprocesses (such as $q\bar q \to b\bar b$) becomes important but are not 
taken into account here.
At small and moderate $E_T$ and $M$, where gluon induced subprocesses play
the leading role, the overall agreement of our predictions
and the data is quite good.
The measured $\Delta \phi$ distribution has a significantly different 
tail at small $\Delta \phi$ compared to the predictions; this will be 
addressed in more detail in the next section.

\subsection{Parton shower effects and additional gluon processes\label{add-effects}} \indent

As shown above, the full set of the data on the heavy flavour production is in general 
reasonably well described. However the measured cross sections as a function of 
$\Delta \phi$ show significant differences. These distributions are directly sensitive 
to additional parton radiation treated by parton showers or by other additional processes.
In the following we investigate the influence of the details of the parton shower and 
the process $gg^* \to g g$ with subsequent branching of one of the outgoing  (off-shell)  $g \to b\bar{b}$ on the $\Delta \phi$ 
distribution. For these studies we use the CCFM set A0. 

\begin{figure}
\centering
\begin{picture}(16.5,8.)(0.,0.)
\put(-0.7,0.){\epsfig{figure=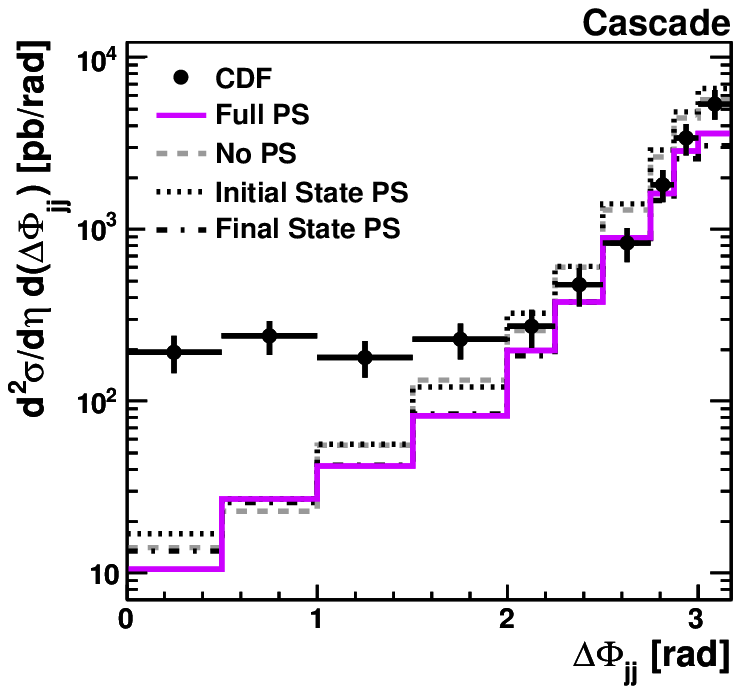, width = 10.4cm}}
\put(7.9,0.){\epsfig{figure=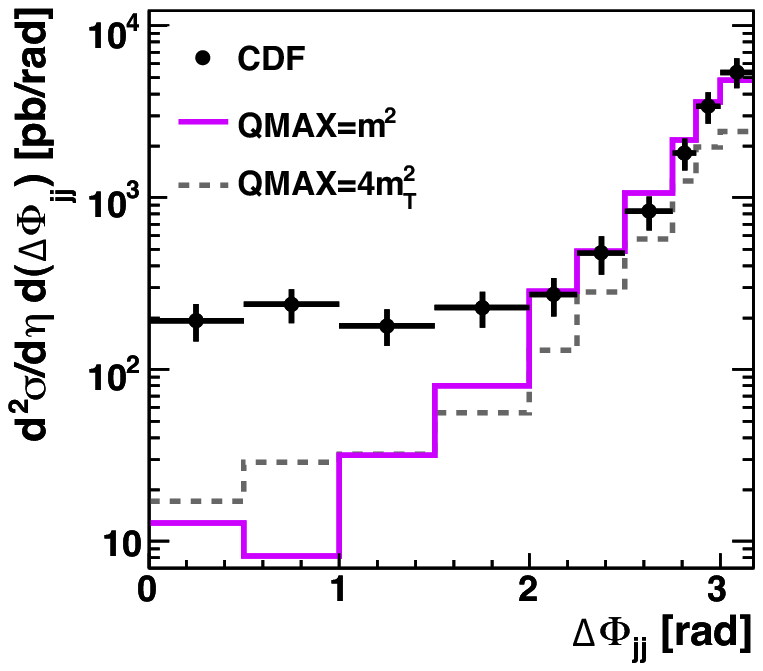, width = 10.4cm}}
\put(4.3,8.){(a)}
\put(12.8,8.){(b)}
\end{picture}
\caption{Parton shower effects for $b$ dijet production in (a). The four lines represent full 
parton shower (solid line), no parton shower (dashed line), initial state
parton shower (dashed dotted line) and final state parton shower (dotted line).
The change of the parton shower scale from $Q_{max}=4m_T^2$ to $Q_{max}=m^2$ 
for $b$ dijet production is depicted in (b). In all cases the CCFM set A0 is used and the data are taken from CDF~\protect\cite{21} and CDF~\protect\cite{6}.}
\label{fig_partonshower}
\end{figure}

In {\sc Cascade}, the initial state parton shower is angular ordered and based 
on the CCFM evolution equations. The final state parton shower is also angular
ordered, but based on the DGLAP evolution equations. To investigate the dependence
on the parton shower, we calculated the cross sections without parton 
shower, only initial state, only final state and both initial and final state parton 
shower, as shown in Fig.~\ref{fig_partonshower}. Here, $b$ dijet production  
is selected as an example. We observe a very small contribution of the initial state parton shower, 
since in $k_T$-factorisation the initial state parton shower does not influence the $k_T$ of 
the gluons (since it is determined from the unintegrated gluon density).
However, a significant
contribution comes from the final state parton shower. The prediction with full parton shower underestimates the back-to-back region 
 of the azimuthal separation of the two $b$ jets.  This suggests 
that the final state parton showering generates too much gluon radiation, which causes 
less correlated jets as shown in Fig.~\ref{fig_partonshower} (a). To investigate this further we changed
the final state parton shower scale $Q^2_{max}$ from $Q^2_{max}=4m_T^2$ to $Q^2_{max}=m^2$. 
As shown in Figure \ref{fig_partonshower} (b), this leads to a higher correlation 
of the $b$ jets and a very good description of the back-to-back region of 
the jets is achieved. 

\begin{figure}
\centering
\begin{picture}(16.5,5.)(0.,0.)
\put(1.,0.8){\epsfig{figure=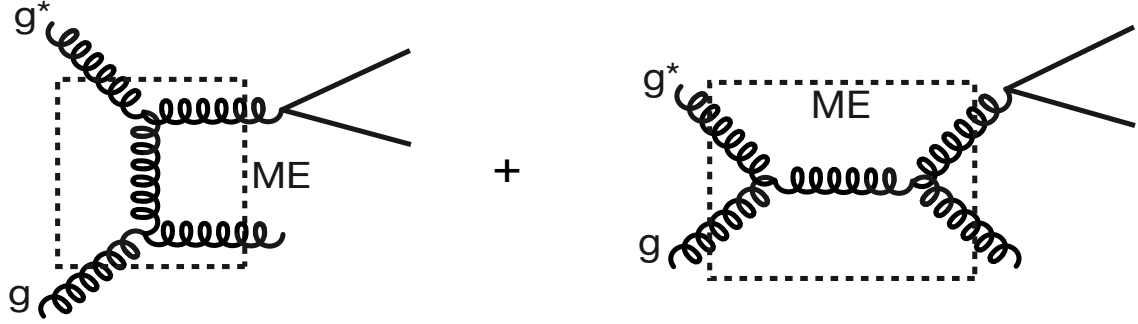, width = 15cm}}
\end{picture}
\caption{The process $gg^*\rightarrow gg$ is shown schematically. The matrix elements are calculated in ~\protect\cite{35}.}
\label{fig_ps_gluon}
\end{figure}

\begin{figure}
\centering
\begin{picture}(16.5,8.)(0.,0.)
\put(-0.7,0.){\epsfig{figure=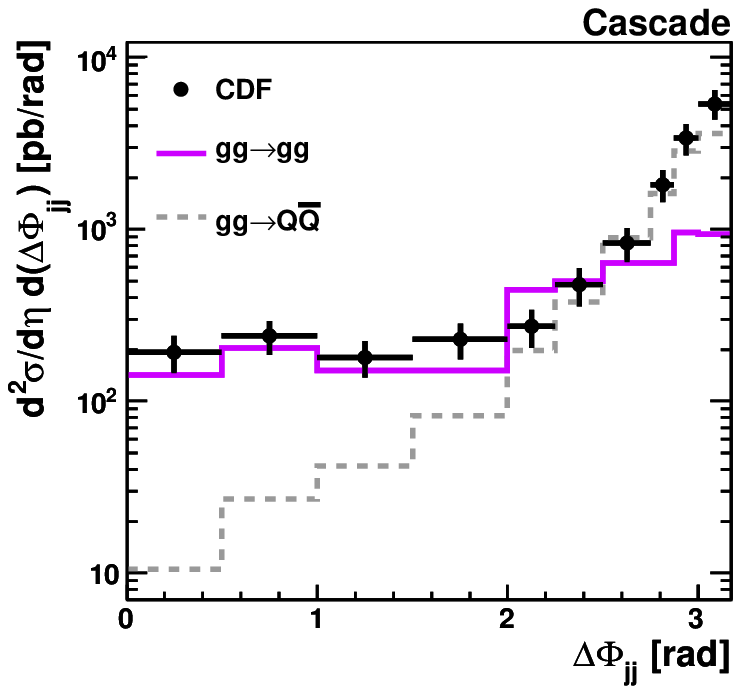, width = 10.4cm}}
\put(7.9,0.){\epsfig{figure=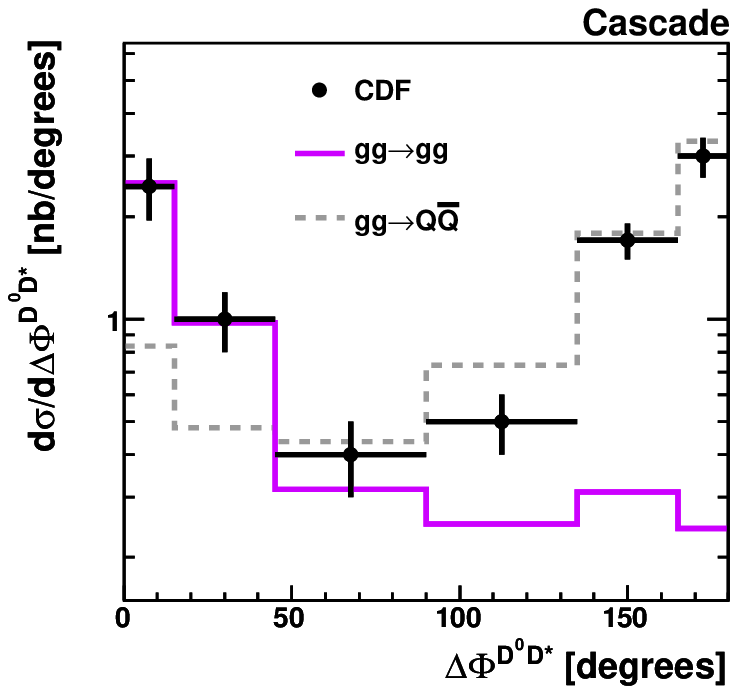, width = 10.4cm}}
\put(4.3,8.){(a)}
\put(12.8,8.){(b)}
\end{picture}
\caption{The processes $g^*g^*\rightarrow Q\bar{Q}$ (dashed line) and $gg^*\rightarrow gg$ for $b$ 
dijet production in (a) and $D^0$, $D^{*-}$ meson production in (b). In all cases the CCFM 
set A0 is used and the data are taken from CDF~\protect\cite{21} and CDF~\protect\cite{6}.}
\label{fig_processes}
\end{figure}

At lower values of the azimuthal separation of the dijet system, higher order processes are  expected to contribute significantly.  Therefore, we repeated the $b$ dijet analysis for the 
process $gg^*\to gg$. In the matrix element calculation, which was performed in~[35], one 
gluon in the initial state is on-shell and the other one is off-shell, as shown in 
Fig.~\ref{fig_ps_gluon}. The heavy quark pair is then produced via parton showers 
in the final state as $g\to b \bar{b}$. As shown in Fig.~\ref{fig_processes} (a), this process contributes significantly to 
the tail of the $\Delta\Phi_{jj}$ distribution very well, without any additional adjustment of parameters. Part of the contribution from $gg^*\to gg$ with $g\to b \bar{b}$ is already included in the matrix element $g^*g^* \to b \bar{b}$, so simply adding both contributions would result in double counting \cite{Serguei}. In \cite{vallecorsa-2007} the influence of multi-parton interactions in the collinear frame is investigated, and it was found, that the contribution even in the small  $\Delta\Phi_{jj}$  region is small, because a large $E_T$ of the jets is required.

In conclusion, the azimuthal cross section measurements can be very well described by adjusting the scale for the final state parton shower to  $Q^2_{max}=m^2$ and by including the process $gg^*\to gg$ with subsequent branching $g\to b\bar{b}$, which in a collinear calculation would contribute only at NNLO.

\section{Conclusions} \indent 

We have studied charm and beauty production 
at Tevatron energies within the framework of the $k_T$-factorization.
Our calculations are based on the CCFM-evolved unintegrated 
gluon densities in a proton.
The analysis covers the total and differential cross sections 
of open charm and beauty quarks, $B$ and $D$ mesons (or rather muons from their 
semileptonic decays). The cross sections of $b\bar b$ di-jet production
have been studied also.
Special attention was put on the specific angular correlations between the 
final-state particles.
Using full hadron-level Monte Carlo generator {\sc Cascade}, we investigated the 
effects coming from the parton showers in initial and final states.
Different sources of theoretical uncertainties have been specially studied.

We obtain good agreement of our calculations
for all observables and the recent
experimental data taken by the D0 and CDF collaborations. 
We have  demonstrated, that the parton shower plays a significant role in the 
description of the cross section.
We obtain a good description of the correlations once the higher order process $gg^* \to gg$ with subsequent $g\to b\bar{b}$ splitting is included.

\section{Acknowledgements} \indent 

We thank S.P.~Baranov for his encouraging interest and useful discussions.
We are very grateful o S. Vallecorsa for many discussions on the CDF results. We are also grateful to S. Baranov for pointing out a possible double counting when $gg\to gg$ processes for heavy quark production.
The authors are very grateful to 
DESY Directorate for the support in the 
framework of Moscow --- DESY project on Monte-Carlo
implementation for HERA --- LHC.
A.V.L. was supported in part by the Helmholtz --- Russia
Joint Research Group.
Also this research was supported by the 
FASI of Russian Federation (grant NS-4142.2010.2) and 
FASI state contract 02.740.11.0244.

\newpage

\newpage

\newpage

\newpage

\newpage

\newpage

\newpage

\newpage

\begin{thebibliography}{35}

\bibitem{1} B.~Abbott {\sl et al.} (D0 Collaboration), Phys. Lett. B {\bf 487}, 264 (2000).
\bibitem{2} B.~Abbott {\sl et al.} (D0 Collaboration), Phys. Rev. Lett. {\bf 84}, 5478 (2000).
\bibitem{3} D.~Acosta {\sl et al.} (CDF Collaboration), Phys. Rev. D {\bf 71}, 092001 (2005).
\bibitem{4} A.~Abulencia {\sl et al.} (CDF Collaboration), Phys. Rev. D {\bf 75}, 012010 (2007).
\bibitem{5} D.~Acosta {\sl et al.} (CDF Collaboration), Phys. Rev. Lett. {\bf 91}, 241804 (2003).
\bibitem{6} J.~Rademacker, Proceedings of Charm'07, Ithaca, NY, August 2007.
\bibitem{7} M.~Cacciari and P.~Nason, Phys. Rev. Lett. {\bf 89}, 122003 (2002);\\
  M.~Cacciari, S.~Frixione, M.L.~Mangano, P.~Nason and G.~Ridolfi, JHEP {\bf 0407}, 033 (2004).
\bibitem{8} M.~Cacciari and P.~Nason, JHEP {\bf 0309}, 006 (2003).
\bibitem{9} B.A.~Kniehl, G.~Kramer, I.~Schienbein, and H.~Spiesberger, Phys. Rev. D {\bf 77}, 014011 (2008).
\bibitem{10} B.A.~Kniehl, G.~Kramer, I.~Schienbein, and H.~Spiesberger, DESY 09-008.
\bibitem{11} S.P.~Baranov and M.~Smizanska, Phys. Rev. D {\bf 62}, 014012 (2000).
\bibitem{12} Ph.~H\"agler, R.~Kirschner, A.~Sch\"afer, L.~Szymanowski and O.V.~Teryaev, Phys. Rev. D {\bf 62}, 071502 (2000).
\bibitem{13} M.G.~Ryskin, Yu.M.~Shabelski and A.G.~Shuvaev, Phys. Atom. Nucl. {\bf 64}, 1995 (2001);\\
  Yu.M.~Shabelski and A.G.~Shuvaev, Phys. Atom. Nucl. {\bf 69}, 314 (2006).
\bibitem{14} N.P.~Zotov, A.V.~Lipatov and V.A.~Saleev, Phys. Atom. Nucl. {\bf 66}, 755 (2003).
\bibitem{15} S.P.~Baranov, N.P.~Zotov and A.V.~Lipatov, Phys. Atom. Nucl. {\bf 67}, 837 (2004).
\bibitem{16} A.V.~Lipatov, L.~L\"onnblad and N.P.~Zotov, JHEP {\bf 0401}, 010 (2004).
\bibitem{17} L.V.~Gribov, E.M.~Levin, and M.G.~Ryskin, Phys. Rep. {\bf 100}, 1 (1983);\\
  E.M.~Levin, M.G.~Ryskin, Yu.M.~Shabelsky and A.G.~Shuvaev, Sov. J. Nucl. Phys. {\bf 53}, 657 (1991);\\
  S.~Catani, M.~Ciafoloni and F.~Hautmann, Nucl. Phys. B {\bf 366}, 135 (1991);\\
  J.C.~Collins and R.K.~Ellis, Nucl. Phys. B {\bf 360}, 3 (1991).
\bibitem{18} E.A.~Kuraev, L.N.~Lipatov and V.S.~Fadin, Sov. Phys. JETP {\bf 44}, 443 (1976);\\
  E.A.~Kuraev, L.N.~Lipatov and V.S.~Fadin, Sov. Phys. JETP {\bf 45}, 199 (1977);\\
  I.I.~Balitsky and L.N.~Lipatov, Sov. J. Nucl. Phys. {\bf 28}, 822 (1978).
\bibitem{19} M.~Ciafaloni, Nucl. Phys. B {\bf 296}, 49 (1988);\\
  S.~Catani, F.~Fiorani and G.~Marchesini, Phys. Lett. B {\bf 234}, 339 (1990);\\
  S.~Catani, F.~Fiorani and G.~Marchesini, Nucl. Phys. B {\bf 336}, 18 (1990);\\
  G.~Marchesini, Nucl. Phys. B {\bf 445}, 49 (1995).
\bibitem{20} B.~Andersson {\sl et al.} (Small-$x$ Collaboration), Eur. Phys. J. C {\bf 25}, 77 (2002);\\
  J.~Andersen {\sl et al.} (Small-$x$ Collaboration), Eur. Phys. J. C {\bf 35}, 67 (2004);\\
  J.~Andersen {\sl et al.} (Small-$x$ Collaboration), Eur. Phys. J. C {\bf 48}, 53 (2006).
\bibitem{21} T.~Aaltonen {\sl et al.} (CDF Collaboration), CDF note 8939.
\bibitem{vallecorsa-2007}
S.~Vallecorsa.
\newblock {\em Measurement of the bbar di-jet cross section at CDF}.
\newblock PhD thesis, University Geneve, 2007.
\bibitem{22} H.~Jung, arXiv:hep-ph/0411287.
\bibitem{23} H.~Jung, Comp. Phys. Comm. {\bf 143}, 100 (2002);\\
  H.~Jung {\sl et al.}, DESY 10-107.
\bibitem{24} F.~Hautmann and H.~Jung, JHEP {\bf 10}, 113 (2008).
\bibitem{25} M.A.~Kimber, A.D.~Martin and M.G.~Ryskin, Phys. Rev. D {\bf 63}, 114027 (2001);\\
  G.~Watt, A.D.~Martin and M.G.~Ryskin, Eur. Phys. J. C {\bf 31}, 73 (2003).
\bibitem{26} M.~Gl\"uck, E.~Reya, and A.~Vogt, Phys. Rev. D {\bf 46}, 1973 (1992);\\
  M.~Gl\"uck, E.~Reya, and A.~Vogt, Z. Phys. C {\bf 67}, 433 (1995). 
\bibitem{27} A.D.~Martin, W.J.~Stirling, R.S.~Thorne, and G.~Watt, Eur. Phys. J. C {\bf 63}, 189 (2009).
\bibitem{28} A.D.~Martin, R.G.~Roberts, W.J.~Stirling, and R.S.~Thorne, Eur. Phys. J. C {\bf 14}, 133 (2000).
\bibitem{29} G.P.~Lepage, J. Comput. Phys. {\bf 27}, 192 (1978).
\bibitem{30} M.~Mangano, P.~Nason and G.~Ridolfi, Nucl. Phys. B {\bf 373}, 295 (1992).
\bibitem{31} C.~Peterson, D.~Schlatter, I.~Schmitt and P.~Zerwas, Phys. Rev. D {\bf 27}, 105 (1983).
\bibitem{32} R.~Barate {\sl et al.} (ALEPH Collaboration), Eur. Phys. J. C {\bf 16}, 597 (2000).
\bibitem{33} K.~Hagiwara {\sl et al.} (PDG Collaboration), Phys. Rev. D {\bf 66}, 010001 (2002).
\bibitem{34} E.~Braaten, K.-M.~Cheng, S.~Fleming and T.C.~Yuan, Phys. Rev. D {\bf 51}, 4819 (1995).
\bibitem{35} M.~Deak, F.~Hautmann, H.~Jung, K.~Kutak, JHEP {\bf 0909}, 121 (2009).
\bibitem{Serguei} S.P.~Baranov, private communication
\end{thebibliography}
\end{document}